\documentclass[journal]{IEEEtran}

\usepackage[pdftex]{graphicx}
\DeclareGraphicsExtensions{.pdf,.jpeg,.jpg,.png,.eps}

\usepackage{multirow}
\usepackage{makecell}
\usepackage{amsmath}
\usepackage{hhline}
\newcolumntype{L}{>{\centering\arraybackslash}m{3cm}}

\usepackage{array}
\usepackage{dblfloatfix}
\usepackage{url}
\usepackage{booktabs}
\usepackage{textcomp}
\usepackage{cite}
\usepackage{hyperref}
\usepackage[table]{xcolor}
\usepackage{lipsum}
\usepackage{subfig}

\makeatletter
\renewcommand*\env@matrix[1][*\c@MaxMatrixCols c]{%
  \hskip -\arraycolsep
  \let\@ifnextchar\new@ifnextchar
  \array{#1}}
\makeatother
\hypersetup{
backref=true
bookmarksnumbered,
pdfstartview={FitH},
citecolor={blue},
linkcolor={blue},
urlcolor={black},
pdfpagemode={UseOutlines}
}

\newlength{\Oldarrayrulewidth}

\begin{document}

\title{A Data Driven End-to-end Approach for In-the-wild Monitoring of Eating Behavior Using Smartwatches}
\author{Konstantinos~Kyritsis,
  \IEEEmembership{Student Member,~IEEE},
  Christos~Diou,~\IEEEmembership{Member,~IEEE},
  and~Anastasios~Delopoulos,~\IEEEmembership{Member,~IEEE}%
  \thanks{K. Kyritsis, C. Diou and A. Delopoulos are with the
    Multimedia Understanding Group, Department of Electrical and
    Computer Engineering, Aristotle University of Thessaloniki,
    Greece.  E-mail: kokirits@mug.ee.auth.gr, diou@mug.ee.auth.gr,
    adelo@eng.auth.gr}%
}


\maketitle

\begin{abstract}
The increased worldwide prevalence of obesity has sparked the interest
of the scientific community towards tools that objectively and
automatically monitor eating behavior. Despite the study of obesity
being in the spotlight, such tools can also be used to study eating
disorders (e.g. anorexia nervosa) or provide a personalized monitoring
platform for patients or athletes. This paper presents a complete
framework towards the automated i) modeling of in-meal eating behavior
and ii) temporal localization of meals, from raw inertial data
collected in-the-wild using commercially available
smartwatches. Initially, we present an end-to-end Neural Network which
detects food intake events (i.e. bites). The proposed network uses
both convolutional and recurrent layers that are trained
simultaneously. Subsequently, we show how the distribution of the
detected bites throughout the day can be used to estimate the start
and end points of meals, using signal processing algorithms. We
perform extensive evaluation on each framework part
individually. Leave-one-subject-out (LOSO) evaluation shows that our
bite detection approach outperforms four state-of-the-art algorithms
towards the detection of bites during the course of a meal (0.923 F1
score). Furthermore, LOSO and held-out set experiments regarding the
estimation of meal start/end points reveal that the proposed approach
outperforms a relevant approach found in the literature (Jaccard Index
of 0.820 and 0.821 for the LOSO and held-out experiments,
respectively). Experiments are performed using our publicly available
FIC and the newly introduced FreeFIC datasets.
\end{abstract}


\section{Introduction}
\IEEEPARstart{R}{eports} published by the World Health Organization
(WHO) formally recognized the global epidemic status of obesity
\cite{world2000obesity}. Regardless of the extensive research that has
been conducted over the years towards understanding obesity, the
reasons behind why people gain excessive weight are still not
completely understood \cite{caballero2007global}. Yet, the positive
energy balance, or in other words when the input energy to one's body
exceeds the output, explains many of the metabolic, endocrine and
behavioral aspects of weight gain
\cite{caballero2007global,world2000obesity}.

Currently, the most common way of objectively and automatically
measuring one's energy expenditure in free-living conditions is done
via physical activity monitoring with the use of accelerometers
\cite{yang2010review,bonomi2012advances}. Even though the field is
still under active research, many solutions have reached product-level
quality, such as Fitbit, Google Fit, Samsung's S Health and Apple's
Health.

At the same time, the \textit{food diary} is still the most common way
of monitoring one's eating habits. Despite being easy to use, the food
diary can suffer in terms of accuracy \cite{schoeller1990accurate},
capacity to measure certain eating behavior parameters that are linked
with obesity such as eating rate \cite{papapanagiotou2018automatic}
and low compliance \cite{korotitsch1999overview}. The lack of tools as
well as the plethora of available sensing devices in the market
motivated the research community to explore automated and objective
monitoring solutions \cite{vu2017wearable,korotitsch1999overview} both
for the in-meal and for the less constrained, in-the-wild measurement
of eating behavior. Besides the study of obesity, there are multiple
domains that can also benefit from the development of automatic and
objective behavior monitoring tools including, but not limited to, the
study of eating disorders (e.g. anorexia nervosa) and the personalized
monitoring for patients or athletes.

Different solutions for measuring in-meal eating behavior have been
proposed in the recent literature. For example, the works of
\cite{zhang2016food,kyritsis2019modeling,dong2012new} use the
information from wrist-mounted inertial sensors to detect food intake
moments. Plate-like weight sensors
\cite{mertes2018measuring,papapanagiotou2018automatic} can estimate
the weight of each individual bite taken throughout the meal. Audio
sensors allow the detection of chewing events
\cite{kalantarian2015audio,passler2012food} and the recognition of
various food types \cite{passler2012food}. More complex sensors like
cameras can recognize food types
\cite{anthimopoulosfood2014,zhu2011multilevel,kawano2013real} and
estimate the caloric intake of a meal by taking pictures of the plate
before and after eating \cite{kong2012dietcam}.

A more challenging problem arises when dealing with in-the-wild
recordings where meals are only a small part of the recorded data. In
such occasions, special care must be taken in order to maintain high
performance while at the same time keep false positives and false
negatives, that can occur during free-living activities, as low as
possible. For example, the works of
\cite{papapanagiotou2017chewing,pabler2011acoustical} use a in-ear
microphone to detect chewing events under free-living
conditions. Inertial information from wrist-mounted sensors can also
be used in order to detect eating
\cite{thomaz2015practical,dong2014detecting} and liquid intake
episodes \cite{gomes2019real}. The work presented in
\cite{ortega2019eating} uses a novel segmentation technique for the
detection and recognition of eating and drinking gestures from
continuous IMU data recorded in free-living conditions. Fusing
inertial and audio information has also shown promising results in the
works of \cite{papapanagiotou2017novel,bi2017familylog}. Additionally,
the work presented in \cite{mirtchouk2017recognizing} showcases how
the combination of free-living and laboratory data with information
from both audio and inertial body-worn sensors can help recognize
eating moments. Eating behavior modeling under free-living conditions
can also be achieved using wearable cameras \cite{zhang2018monitoring}
and jaw motion sensors \cite{fontana2014automatic}. Finally, the
authors of \cite{mirtchouk2019automated} combine the information from
body-worn audio and inertial sensors and use a semi-supervised
hierarchical classification scheme to identify the food type in
unrestricted environments.

In this paper we propose a complete framework for automatically
measuring eating behavior from in-the-wild collected inertial data
(acceleration and orientation velocity) using a smartwatch. Initially,
we show how a purely data-driven approach that uses an end-to-end
\textit{Neural Network} (NN) with both convolutional and recurrent
layers can be used to detect food intake events (i.e. bites). Then we
present how the distribution of detected bites throughout the day can
be used to isolate meal events and detect their start and end
time-points. In our experimental section we extensively evaluate each
part of the framework individually, in terms of in-meal bite detection
and in terms of detecting meal end-points under free-living
conditions, and compare it with other state-of-the-art approaches
found in the literature. In all experiments we use our \textit{Food
  Intake Cycle} (FIC) and the newly introduced FreeFIC datasets. Both
datasets are publicly available on the Multimedia Understanding Group
webpage\footnote{Available at:
  \url{https://mug.ee.auth.gr/intake-cycle-detection}}.

The rest of the paper is organized as follows. Section \ref{sec:soa}
provides a review of the relevant work regarding: i) the in-meal
detection of eating moments and ii) the localization of eating events
throughout the day using in-the-wild collected data. Sections
\ref{sec:algorithm} and \ref{sec:inthewild} describe the steps of the
proposed data driven bite detection and meal localization
approaches. Section \ref{sec:experimentsevaluation} presents the
datasets, the experiments and results. Section \ref{sec:limitations}
presents the limitations of the proposed approach. Finally, Section
\ref{sec:conclusions} concludes the paper.

\section{Related work}
\label{sec:soa}
A number of works have been proposed in the recent literature
\cite{heydarian2019assessing} for measuring in-meal eating behavior by
detecting food intake events using inertial data from wrist mounted
sensors. The common denominator between such works is that they aim at
recognizing specific wrist movements (or series of wrist movements)
that are part of the process of delivering food to the mouth.

The authors of \cite{dong2012new} use a single gyroscope channel to
detect a characteristic wrist-roll motion that happens during a bite
event. The same algorithm was used to produce the results presented in
\cite{shen2017assessing}. Specifically, a bite is detected as a
sequence of four different conditions. The first two conditions have
to do with the velocity initially surpassing a positive and then a
negative threshold. The final two conditions deal with the minimum
amount of time that needs to pass between the two aforementioned roll
motions and between two consecutive detected bites. The authors of
\cite{shen2017assessing} report a sensitivity/positive predictive
value of $0.81$/$0.86$ using data collected in the lab and
$0.81$/$0.83$ using more realistic data collected in the cafeteria.

The most recent work of our group \cite{kyritsis2019modeling},
proposes an in-meal bite detection method that uses the inertial data
($3$D acceleration and orientation velocity) from off-the-shelf
smartwatches. In more detail, we use five specific wrist
\textit{micromovements} (namely pick food, upwards, downwards, mouth
and no movement) to model the series of actions leading to and
following an intake event. The method operates in two steps. In the
first step we process windows of raw sensor data to estimate the
micromovement probability distribution using a Convolutional Neural
Network (CNN). During the second step, we use a Long-short Term Memory
(LSTM) network to capture the temporal evolution of micromovements and
classify sequences as food intakes cycles. Leave-one-subject-out
(LOSO) evaluation results on our public FIC dataset of $21$ meals from
$12$ subjects yield a promising F$1$ score of $0.913$. Despite the
satisfying performance, annotating micromovements in IMU sequences
using visual information is an expensive and error-prone process.

An approach that revolves around a similar micromovement-based concept
as \cite{kyritsis2019modeling} is presented by Zhang \textit{et al.}
in \cite{zhang2016food}. The authors propose the use of two wrist
gestures in order to characterize a motion as a feeding gesture,
namely: ``food to mouth'' and ``back to rest''. Their processing
pipeline starts by preprocessing the acceleration and orientation
velocity streams, continues with the extraction of $66$ statistical
features using a sliding window approach and a classification scheme
to characterize windows as feeding or non-feeding gestures. Finally,
bite detection is achieved by clustering the detected feeding gestures
using \textit{Density-Based Spatial Clustering of Applications with
  Noise} (DBSCAN). The authors evaluate their algorithm using a
leave-one-subject-out cross validation scheme on their dataset of
$15$ subjects. In their experimental section, the authors report an
F$1$ score of $0.757$ using the AdaBoost classifier and the most
descriptive feature subset.

In another work of our group \cite{kyritsis2018end} we present a
data-driven approach towards the detection of bites during the course
of a meal without using the additional knowledge of
micromovements. The core of our approach is an \textit{Artificial
  Neural Network} (ANN) with convolutional and recurrent
layers. Experimental results with a smaller version of the FIC dataset
($10$ meals from $10$ subjects) showed that the data-driven method
achieved similar performance as a micromovement-based approach
\cite{kyritsis2017food}, with F$1$ scores of $0.884$ and $0.892$
respectively. In the work presented in this paper we further expand
the works of \cite{kyritsis2017food,kyritsis2018end} by increasing the
knowledge base of the model and moving further way from a
micromovement-based approach. This allows the convolutional part of
the network to discover the optimal features representations before
performing temporal modeling using an LSTM.


The approaches discussed so far focus on measuring and modeling
\emph{in-meal} eating behavior i.e., behavior during the course of a
meal. On the other hand, a much smaller body of research aims at
identifying and localizing eating episodes (such as meals or snacks)
from data collected using wrist-mounted inertial sensors.

In \cite{fontana2014automatic} the authors make use of a novel sensing
platform that incorporates a jaw motion sensor, a hand gesture sensor
and an accelerometer towards food intake monitoring in free-living
environments. After a preprocessing step, the authors follow a feature
extraction scheme on the data of each sensor before performing early
fusion using an relatively shallow \textit{Artificial Neural Network}
(ANN) consisted by an input, a single hidden and an output
layer. Results on their dataset of $12$ subjects wearing the sensing
platform for $24$ hours reveal that the system is able to detect food
intakes with an accuracy of $0.898$. The same dataset of $12$ people
was also used in \cite{farooq2016detection} as a benchmark for three
different ensemble techniques, boosting, bootstrap aggregation and
stacking, trained with three different weak classifiers: i) Decision
Trees, ii) Linear Discriminant Analysis and iii) Logistic
Regression. Following the same feature extraction scheme as in
\cite{fontana2014automatic} the authors of \cite{farooq2016detection}
show that using bootstrap aggregation with Fisher LDA as the base
classifier leads to an overall improvement of $0.04$ ($0.938$ accuracy
instead of $0.898$). In both works however,
\cite{fontana2014automatic} and \cite{farooq2016detection}, the
authors do not provide a final estimate of the meal start and end
points; instead they classify segments as food intake or not.

The work of \cite{bi2017familylog} uses a combination of wrist-mounted
sensors and smartphones able to capture the wrist motion and the
surrounding sound of the involved participants with the purpose of
detecting family mealtime activities. By fusing the audio and motion
data with a \textit{Hidden Markov Model} (HMM), the authors of
\cite{bi2017familylog} achieve an average precision and recall of
$0.807$ and $0.895$ regarding the detection of family meals.

Dong \textit{et al.} in their work \cite{dong2014detecting} use a
conventional smartphone strapped on the participant's wrist to capture
the acceleration and orientation velocity signals throughout the day
in uncontrolled environments. Their work is based on the hypothesis
that a period of increased wrist motion energy exists before and after
every meal, while during the meal the wrist motion energy is
decreased. In more detail, the authors use a heuristic peak detector
based on the concept of hysteresis threshold to create potential meal
segments and then follow a feature extraction scheme and a Naive Bayes
classification scheme to classify the segmented sequences as eating or
not. In their large dataset of $43$ subjects the authors yield a
sensitivity of $0.81$, a specificity of $0.82$ and a weighted accuracy
(with a true positive to true negative ratio of $20$:$1$, due to the
limited time spend eating throughout the day) of $0.82$. The later
work of \cite{sharma2016automatic} extended the work of
\cite{dong2014detecting} by using a more suited novel smartwatch-like
sensing platform and increasing the dataset's size by over twofold,
reaching a total of $104$ subjects (up from $43$). Evaluation results
on the larger dataset using the same meal detection method as in
\cite{dong2014detecting} indicate a lower performance by yielding a
sensitivity of $0.69$ (from $0.81$), a specificity of $0.80$ (from
$0.82$) and a weighted accuracy of $0.75$ (from $0.82$). In both cases
(\cite{dong2014detecting} and \cite{sharma2016automatic}) the authors
suggest that their initial hypothesis (i.e. increased wrist activity
before and after a meal) may not work for all participants.

In our work presented in \cite{kyritsis2019detecting} we show how the
distribution of bite detections, produced by the data-driven method of
\cite{kyritsis2018end}, throughout the day can be used to effectively
detect the meal start and end points from in-the-wild collected
data. Our meal localization method is based on the hypothesis that the
density of detected bites is high during meals and low when outside of
meals.

In the current work we present a complete framework that is capable of
both estimating the in-meal eating behavior and temporally localizing
eating episodes during the day from in-the-wild data. Specifically, on
the first part of this paper we show how an end-to-end NN that uses
both convolutional and recurrent layers, which are jointly trained
with the same loss function, can be used for detecting food intake
moments during a meal given the raw inertial series. On the second
part we present how we can use the distribution of bite detections as
produced by the end-to-end NN in order to localize eating episodes
during the day, using signal processing algorithms.

Briefly, the main contributions of our current work are:

\begin{itemize}
\item A complete framework that includes:
  \begin{itemize}
  \item A novel preprocessing step for adjusting the orientation of
    the IMU frames that belong to the opposite wrist than the one used
    as reference; an important step to ensure data uniformity prior to
    any learning process (Section \ref{sec:adjust}).
  \item A data-driven approach for detecting food intake events
    (i.e. bites) during the course of a meal using an end-to-end NN
    with both convolutional and recurrent layers (Section \ref{sec:algorithm}).
  \item An algorithm for the temporal localization of eating episodes
    using the distribution of bite detections produced by the
    end-to-end NN (Section \ref{sec:inthewild}).
    \end{itemize}
\item Two publicly available datasets of in-the-wild collected data
  that contain a wide spectrum of unscripted every-day
  activities. Both datasets contain high rate $3$D accelerometer and
  gyroscope measurements originating from commercial smartwatches. The
  FreeFIC dataset (Section \ref{sec:freefic_data}) contains a total of
  $16$ in-the-wild recordings belonging to $6$ unique subjects with a
  total duration of $77.32$ hours. The FreeFIC held-out dataset
  (Section \ref{sec:freefic_heldout_data}) contains $6$ in-the-wild
  recordings from $6$ unique subjects with a total duration of $35.39$
  hours. As annotations we use the start and end moment of meals as
  self-reported by the participants. At the time of writing, FreeFIC
  and FreeFIC held-out are the only datasets that contain high-rate
  information from a single smartwatch and in-the-wild meal sessions
  that make use of the fork, spoon and knife that are available to the
  public.
\item An extensive experimental evaluation of both the proposed
  in-meal bite detection method against four other state of the
  art-methods (Section \ref{sec:exi}) and the proposed meal
  localization method against the only similar approach found in the
  literature (Sections \ref{sec:exii} and \ref{sec:exiii}).
  \begin{itemize}
  \item The proposed meal localization method's ability to generalize
    on previously unseen, in-the-wild, data is also further validated
    using an external publicly-available dataset (ACE Free-living by
    Mirtchouk \textit{et al.} presented in
    \cite{mirtchouk2017recognizing}) that contains \textit{similar}
    information (Section \ref{sec:exiv}). The ACE Free-living dataset
    contains meal recordings collected in-the-wild; however, it should
    be emphasized that has properties that differ (in terms of
    sampling rate, number of smartwatches, use of utensils and meal
    types) from the FreeFIC/FreeFIC held-out datasets.
  \end{itemize}
 \item In our experiments we also showcase the positive effects of a
   novel augmentation scheme for inertial data that can increase the
   F$1$ score regarding the in-meal detection of bites from $0.888$ to
   $0.923$ (Section \ref{sec:exi}).
\end{itemize}

\section{End-to-end detection of bites}
\label{sec:algorithm}
We use the term \textit{end-to-end} to describe a learning mechanism
with the ability of extracting problem-specific data representations
from the raw input and subsequently model the evolution of those
representations across time. In the context of our work, the
end-to-end learning mechanism is a recurrent ANN. The proposed
approach uses a CNN to handle the extraction of features, followed by
an LSTM network to model their temporal evolution. Both parts of the
ANN are jointly trained by minimizing a single loss function using
backpropagation. This approach differs significantly from our previous
two-step work \cite{kyritsis2019modeling} where we made use of the
explicit knowledge of hand micromovements and food intakes to
\textit{separately} train a CNN and an LSTM network.

In this section we present a method for processing the raw triaxial
acceleration and orientation velocity signals that originate from a
commercial smartwatch with the aim of detecting bite events. Following
the preprocessing step, the ANN is trained in an end-to-end fashion
solely using the information of food intake moments as ground
truth. In addition, on-line inference, i.e. without the need for
presegmenting sequences, enables us to fully take advantage of LSTM's
memory capabilities. The bite moments are finally detected using
signal processing algorithms on the ANN prediction signal. The
following subsections provide details on the proposed bite detection
pipeline.
\subsection{Preprocessing}
\label{sec:preproc}
In this work we make use of the $6$ degrees of freedom (DoF) inertial
data from a commercial smartwatch. Specifically the $x$, $y$ and $z$
streams of the accelerometer and gyroscope sensors denoted as
$\mathbf{a}_x, \mathbf{a}_y, \mathbf{a}_z, \mathbf{g}_x, \mathbf{g}_y$
and $\mathbf{g}_z$. For a single point in time $m$ the vector
$\mathbf{x}(m)=[a_{x}(m), a_{y}(m), a_{z}(m),g_{x}(m), g_{y}(m),
  g_{z}(m)]^\top$ contains the instantaneous acceleration and
orientation velocity. A complete recording with a duration of
$t_{tot}$ seconds and a sensor sampling frequency of $f_{s}$
\textit{Hz} can be represented as $\mathbf{R} = [\mathbf{x}(1),
  \ldots, \mathbf{x}(M)]^\top$, where $M=t_{tot} \cdot f_s$ is the
length of the recording in samples.
\subsubsection{Hand mirroring}
\label{sec:adjust}
\begin{figure}[t]
  \centering
  \includegraphics[width=1\linewidth]{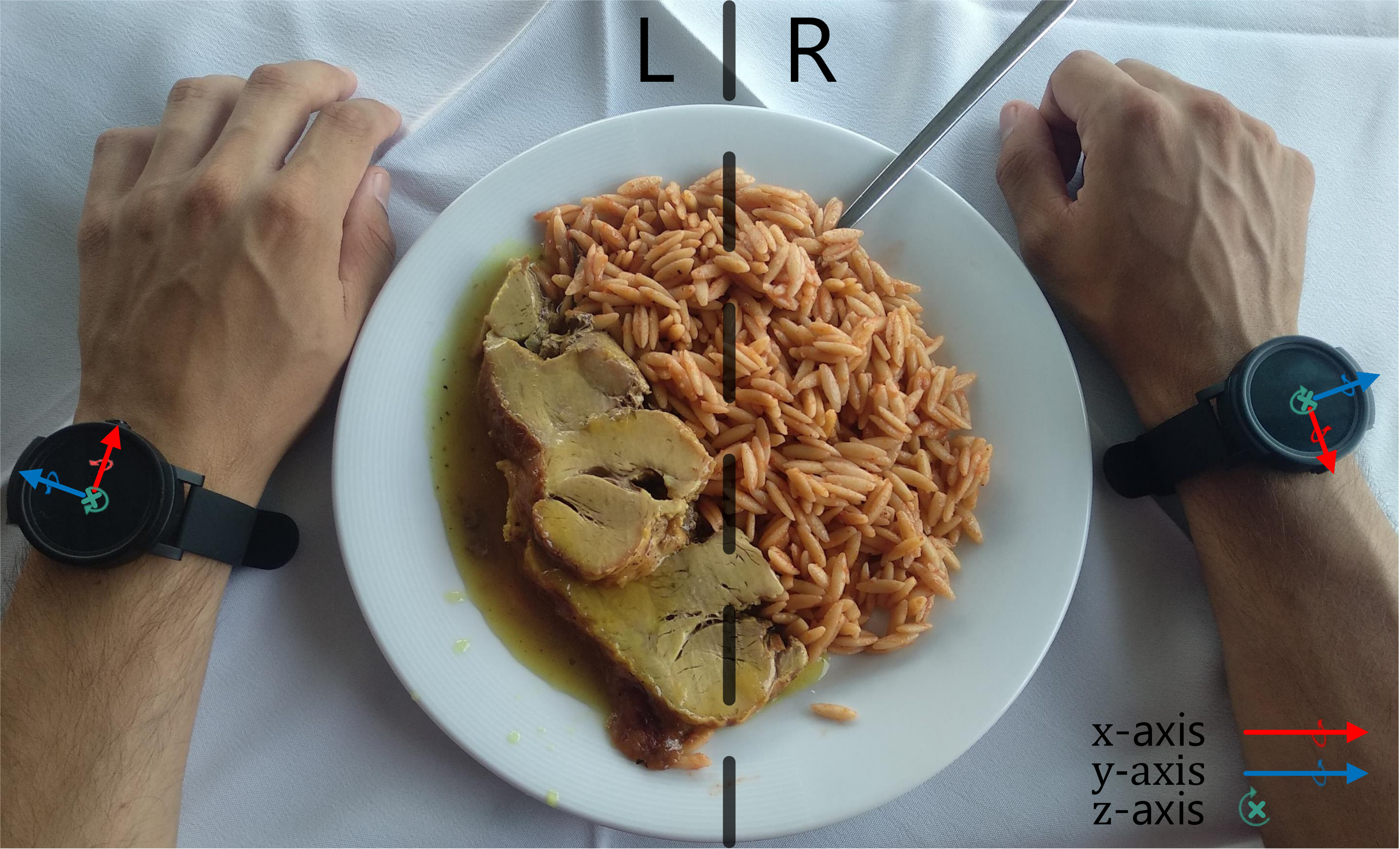}
  \caption{Figure depicting the orientation of the IMU sensors when
    the smartwatch is worn on either the left (L) or the right (R)
    hand of a participant. The figure also illustrates the need for
    adjusting the orientation of the accelerometer's $x$ and
    gyroscope's $y$ and $z$ axes in a way that when a hand moves then
    the movements of the opposing hand are reflected as if a mirror
    was placed on the dotted line. In all cases, the direction of the
    gyroscope's axes is given by the \textit{right hand rule}.}
  \label{fig:hands}
\end{figure}

It is expected from people to eat their meals using either of their
hands to operate the spoon and/or fork. As a result, properly
adjusting the IMU frames to a common reference is an important
preprocessing step prior to any learning process. In our work we
selected the participant's \textit{right} hand as the reference. We
achieve orientation adjustment by appropriately transforming all
left-handed recordings $\mathbf{R}_{l}$ by changing the
\textit{direction} of the $\mathbf{a}_{x}$, $\mathbf{g}_{y}$ and
$\mathbf{g}_{z}$ sensor streams. We define this process as
\textit{hand mirroring}. Formally, this is depicted in Equation
\ref{eq:rot}.
\begin{equation}
\label{eq:rot}
\mathbf{\widetilde{R}}_{r} = \mathbf{R}_{l} \times \begin{bmatrix} -1 & 0
  & 0 & 0 & 0 & 0 \\ 0 & 1 & 0 & 0 & 0 & 0 \\ 0 & 0 & 1 & 0 & 0 & 0
  \\ 0 & 0 & 0 & 1 & 0 & 0 \\ 0 & 0 & 0 & 0 & -1 & 0 \\ 0 & 0 & 0 & 0
  & 0 & -1 \end{bmatrix}
\end{equation}
Where $\mathbf{R}_l$ is the initial left-handed recording and
$\mathbf{\widetilde{R}}_r$ the mirrored, both with dimensions $M\times
6$. All original right-handed recordings $\mathbf{R}_r$ are left
unprocessed. Figure \ref{fig:hands} shows how the inertial sensors are
orientated when the smartwatch is either worn on the left or on the
right wrist of the participant. A closer look in Figure
\ref{fig:hands} will initially reveal that no adjustment is required
for the $y$ and $z$ axes of the accelerometer as they are already
mirrored. However, the direction of the $x$ accelerometer axis needs
to be adjusted in such a way that e.g., a movement of both wrists
along the $x$ axis \textit{towards} the torso of the participant would
yield the same measurements. In the previous example and given the
setup presented in Figure \ref{fig:hands} the $x$ accelerometer stream
of the right wrist would yield positive measurements while the left
wrist would yield negative. The same is also true for the gyroscope's
$y$ and $z$ axes. For example a rotation of both wrists around the $y$
axis towards the participants torso would again yield positive
velocity measurements for the right wrist and negative for the
left. Finally, a rotation of the wrists around the $z$ axis towards
the dotted line would yield negative velocity measurements for the
right wrist and positive for the left one. In all cases presented
above, the transformation presented in Equation \ref{eq:rot}
appropriately adjusts the orientation so the left wrist movements
mirror the ones from the right wrist.

\subsubsection{Smoothing \& gravity removal}
\label{sec:signalpreproc}
To deal with the small fluctuations introduced by the sensors'
hardware we individually convolve each of the $3$D accelerometer and
gyroscope streams with a moving average filter. After experimenting
with different filter lengths we selected a length of $l_{ma}=25$
(considering a sampling frequency $f_s=100$ Hz) as it produced
satisfactory results with minimal distortion. Each tap weight of the
moving average filter was set to $1/25$.

Furthermore, since the accelerometer sensor captures the acceleration
caused due to the Earth's gravitational field in addition to the
voluntary wrist movements, the final preprocessing step is to
attenuate the Earth component. We achieve this by individually
convolving the $\mathbf{a}_x$, $\mathbf{a}_y$ and $\mathbf{a}_z$
components of $\mathbf{R}$ with a high-pass finite impulse response
(FIR) filter. The filter's cutoff frequency $f_c$ and length of the
tap delay line $l_{hp}$, were set to $1$ \textit{Hz} and $512$ samples
respectively.

\subsection{End-to-end network architecture}
\label{sec:end2end_arch}

The proposed end-to-end architecture is comprised of two parts, the
first one being the convolutional and the second the recurrent. The
convolutional part has a depth of three $1$D convolutional layers with
the first two being followed by max pooling operations that decimate
the activations of the previous layer by a factor of $2$. The number
of filters used in each convolutional layer increases as the depth
increases. Specifically, we used $32$, $64$ and $128$ as the number of
filters in each of the three convolutional layers. The filter lengths
for each layer were selected to be $5$, $3$ and $3$ samples
respectively; or $\frac{5}{fs}$, $\frac{3}{fs/2}$ and $\frac{3}{fs/4}$
\textit{secs} due to the intermediate max pooling operations. All
convolutional layers use the Rectified Linear Unit (ReLU) as the
non-linearity for their activations, defined as $\rho(x)=\max(0,x)$.

Overall, the convolutional part of our network is inspired by the
popular VGG \cite{simonyan2014very} architecture for image
classification; essentially increasing the number of filters in the
convolutional layers after each max pooling application while keeping
the same kernel sizes. Contrary to VGG, however, our architecture uses
one convolution layer before max pooling instead of using pairs of
convolution layers in-between max pooling layers. This helps in
keeping the number of model parameters low.

\begin{figure*}[t]
  
  \includegraphics[width=1.0\linewidth]{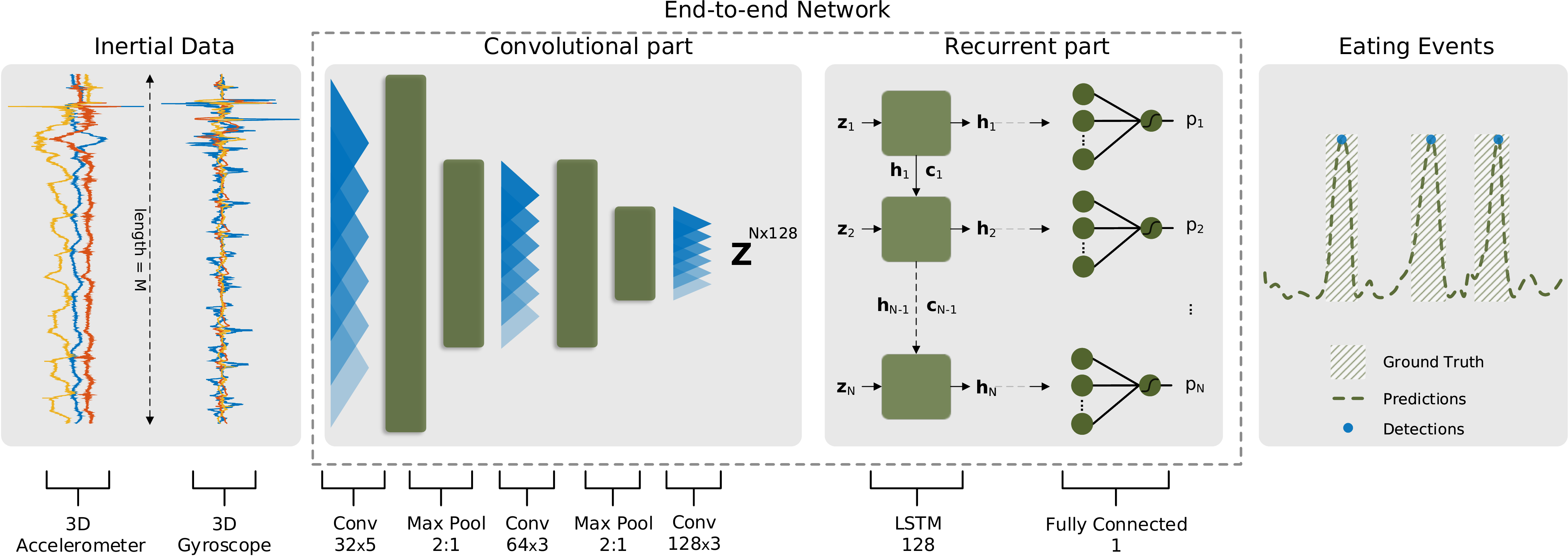}
  \caption{Overall pipeline of the in-meal bite detection
    approach. From left to right, the preprocessed inertial signals
    ($3$D acceleration and orientation velocity) of length $M$ that
    are captured during the meal are forwarded to the end-to-end
    network. The network processes the $M\times 6$ data, (essentially
    $\mathbf{R}$ from Section \ref{sec:preproc}), and outputs the
    $N$-dimensional bite probability vector $\mathbf{p}$, where
    $N=M/4$ due to the two max pooling operations. Variable
    $\mathbf{Z}$ with dimensions $N\times128$ represents the
    intermediate output of the convolutional part of the end-to-end
    network while $h_i$ and $c_i$ represent the $i$-th output and cell
    state of the LSTM block.}
  \label{fig:arch}
\end{figure*}

The recurrent part consists of a single LSTM layer with $128$ hidden
cells. We used the hard sigmoid function, defined as $\sigma_{h}(x) =
\max(0, \min(1, 0.2 \cdot x +0.5))$, as the activation of the LSTM's
recurrent steps. During training the output of the network is obtained
by propagating the \textit{final} $128$-dimensional output of the LSTM
layer to a fully connected layer with a single neuron and the sigmoid
function as the activation, defined as $\sigma(x) =
(1+e^{-x})^{-1}$. The total number of learnable parameters for the
proposed network is $163,617$.

On the other hand, during inference we modified the LSTM layer to
provide all intermediate outputs to the fully connected layer and
subsequently the fully connected layer to provide an output $p$ for
all intermediate outputs of the LSTM layer. In more detail, given a
$M\times 6$ meal recording $\mathbf{R}$, the convolutional part of the
inference network generates the $N\times128$ \textit{internal}
variable $\mathbf{Z}$, with $N=M/4$ due to the two max pooling
operations that decimate the activations by a factor of $2$
each. Subsequently, the recurrent part of the network (i.e. the LSTM
layer) processes $\mathbf{Z}$ to produce the final $N\times 1$
prediction vector $\mathbf{p}$. Figure \ref{fig:arch} depicts the
architecture of the inference network as part of the in-meal bite
detection pipeline.

\subsection{Training the network}
\label{sec:end2end}
For training examples, we used a sliding window of length $w_l$ and
step $w_s$ to extract data frames from all meals in the training set.
As a result, we obtain the set
$\mathcal{X}=\{\mathbf{X}_{1},\mathbf{X}_{2},\cdots\}$, that contains
the training examples, $\mathbf{X}_{i}$ each with dimensions $w_l
\times 6$, from all meals. The process of extracting training samples
continues by pairing each $\mathbf{X}_i$ with a label $y_{i}$. A label
$y_{i}$ is positive when the \textit{end} of the sliding window is
within a distance ($\pm \epsilon$) from the \textit{end} of a intake
cycle event and negative for every other case. In the evaluation
section (Section \ref{sec:dataset}) we present two different
approaches for assigning values to $y_i$ depending on the available
ground truth data. It should also be noted that during training the
internal variable $\mathbf{Z}$ has dimensions
$\frac{w_l}{4}\times128$.

We also augmented the training set by artificially changing the
orientation of the smartwatch on the wrist. In more detail, each
example in the minibatch has a $50\%$ chance to be selected for
transformation. If the example is selected then two random numbers,
$\hat{\theta}_{x}$ and $\hat{\theta}_{z}$, are drawn from a normal
distribution with mean and standard deviation equal to $0$ and $10$
respectively. 

Then based on $\hat{\theta}_{x}$ and $\hat{\theta}_{z}$ we create two
$3\times3$ rotation matrices, namely $\mathbf{Q}_x(\hat{\theta}_{x})$
and $\mathbf{Q}_z(\hat{\theta}_{z})$ that correspond to rotation
around the $x$ and $z$ axes, respectively. The final transformation
$\mathbf{Q}$ is randomly chosen to be one of the following: i)
$\mathbf{Q}_x(\hat{\theta}_{x})$, ii)
$\mathbf{Q}_z(\hat{\theta}_{z})$, iii) $\mathbf{Q}_x(\hat{\theta}_{x})
\cdot \mathbf{Q}_z(\hat{\theta}_{z})$ or iv)
$\mathbf{Q}_z(\hat{\theta}_{z}) \cdot
\mathbf{Q}_x(\hat{\theta}_{x})$. The transformation is then applied to
the chosen $i$-th training example $\mathbf{X}_{i}$ in the batch as:

\begin{equation}
  \label{eq:transform_rot}
 \mathbf{X}'_{i} = \left( \left[ \begin{array}{@{\,} c|c @{\,}}
  \mathbf{Q} & \mathbf{0}_{3 \times 3}\\      
  \hline
  \mathbf{0}_{3 \times 3} & \mathbf{Q} \\
    \end{array} \right] \cdot \mathbf{X}_{i}^{\top}\right)^{\top}
\end{equation}

\noindent Where $\mathbf{0}_{3 \times 3}$ is used to represent a
$3\times 3$ matrix of zeros. We selected not to perform any
augmentation by rotating the $\mathbf{X}_{i}$ samples around the $y$
axis (i.e. by $\mathbf{Q}_y(\hat{\theta}_{y})$) as such rotation of
the smartwatch on the wrist is not physically feasible. Applying such
transformations allows us to mimic different positions
($\pm10^{\circ}$) of the user's smartwatch around the wrist. This
allows us to introduce variability in the examples provided to the
network during training.

During training the network minimizes the binary cross entropy loss,
\begin{equation}
  \label{eq:bincross}
  \mathcal{{L}} = - \sum\limits_{i\in K} \Big( \hat{y}_{i}
  \log(\hat{p}_{i}) - (1-\hat{y}_{i}) \log(1-\hat{p}_{i}) \Big)
\end{equation}
where $\hat{y}\in\{-1,+1\}$ is the target, $\hat{p}_{i}$ is the
network's prediction for the $i$-th sequence in a mini-batch that
contains $K$ examples. We also applied a $50\%$ dropout chance to the
inputs of the fully connected layer as a form of regularization in
order to avoid overfitting during training
\cite{srivastava2014dropout}. Finally, the network is trained using
the RMSProp optimizer with a learning rate of $10^{-3}$ and a batch
size of $128$ samples. Each batch contained an equal number of
positive and negative training examples to avoid bias. After
inspecting the learning curves regarding the training accuracy and
cross entropy loss on the training set we observed only a marginal
improvement after the $5$th epoch; therefore, we selected $5$ as the
number of epochs for training the network.

\subsection{Bite detection}
\label{sec:bitedet}
By processing an $M\times 6$ recording $\mathbf{R}$ that contains the
preprocessed inertial observations, the inference end-to-end network
outputs the $N$-length prediction series $\mathbf{p}$, with $N=M/4$
due to the max pooling operations (Section \ref{sec:end2end_arch}). We
perform bite detection by initially replacing with zeros the elements
of $\mathbf{p}$ that are lower than a probability threshold
$\lambda_{p}$. Next, by performing a local maxima search, with a
minimum distance between two consecutive peaks set at $2$ seconds, on
the thresholded series $\mathbf{p}$ we obtain the set of detected
bites $\mathcal{B}=\{b_{1},\ldots,b_{L}\}$. Each $b_{i}$ element is
essentially the timestamp that corresponds to the $i$-th peak.

Experimenting with a small part of the FIC dataset led us to select
$0.89$ for the $\lambda_{p}$ threshold, as this value yielded the
highest F$1$ score ($0.932$). This set included $4$ out of the $21$
available meals (approximately $19.04\%$), selected at random, and
$319$ ground truth bites out of the total $1,332$ (approximately
$23.94\%$).

\begin{figure}[t]
  \centering
  \begin{tabular}{@{}c@{}}
    \includegraphics[width=1\linewidth]{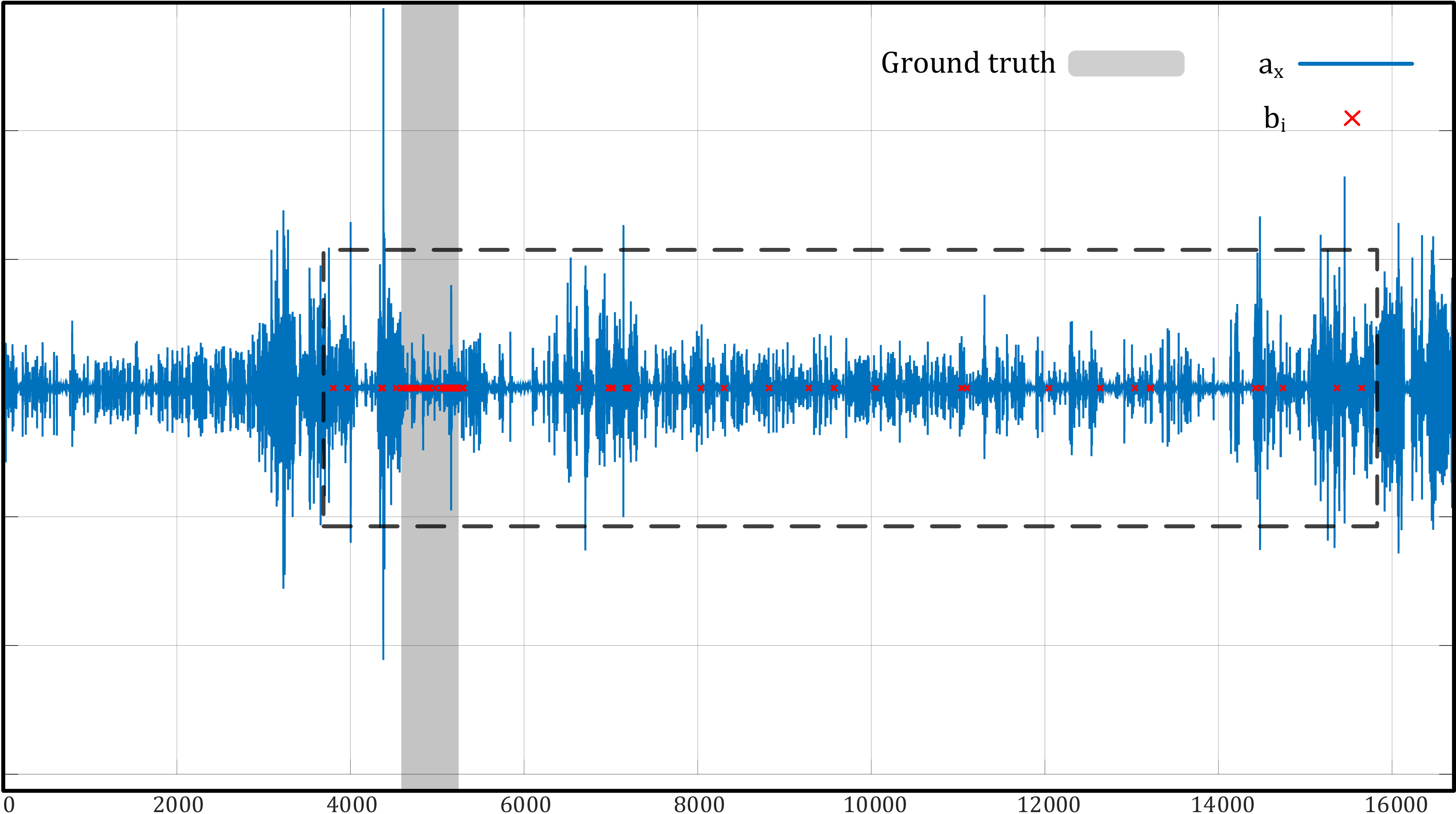} \\[\abovecaptionskip]
    \small (a) 
  \end{tabular}

  \vspace{\floatsep}

  \begin{tabular}{@{}c@{}}
    \includegraphics[width=1\linewidth]{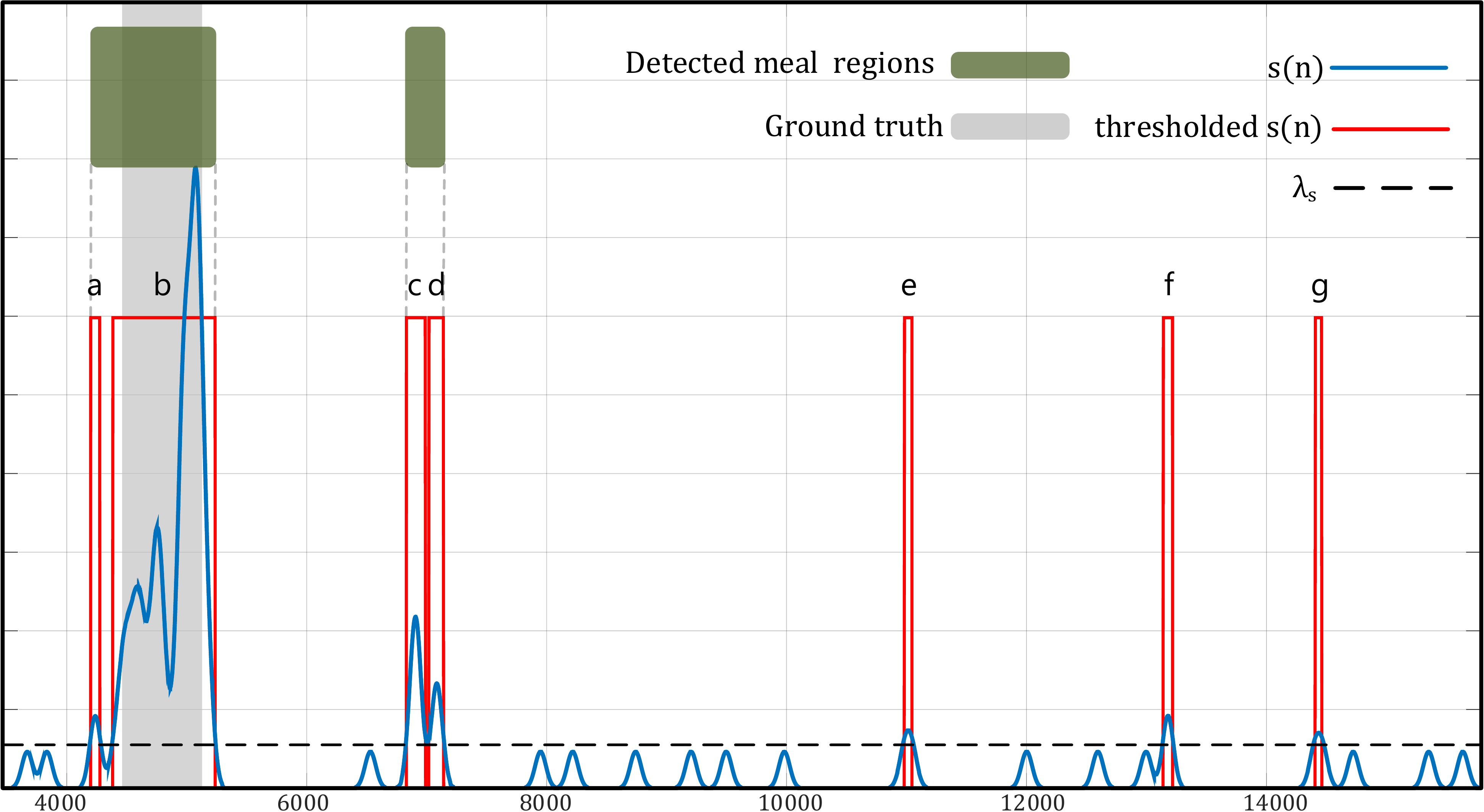} \\[\abovecaptionskip]
    \small (b) 
  \end{tabular}

  \caption{Example estimation of meal end-points. In both sub-figures
    the ground truth is depicted with the shaded background and the
    horizontal axis represents the time in seconds. Specifically,
    sub-figure a) shows the $\mathbf{a}_x$ axis of the accelerometer
    sensor and each red $\times$ symbol represents a bite detection
    $b_i$. Sub-figure b) focuses on the area of a) that is marked with
    the dotted line. It depicts the smoothed series $s(n)$ and how
    thresholding it by using $\lambda_{s}$ as a cutoff point we obtain
    the candidate meal intervals denoted with a, b, c, d, e, f and
    g. By iteratively merging nearby intervals and subsequently
    rejecting short meals, we estimate the final meal regions. In this
    specific example, intervals e, f and g where rejected while a and
    b as well as c and d were respectively merged and formed the two
    final meal estimates (depicted with green
    background).}\label{fig:meal_figs}
\end{figure}
\section{In-the-wild meal detection}
\label{sec:inthewild}
The term \textit{in-the-wild} is usually used to describe the
unconstrained nature of capturing conditions. In the context of this
work, in-the-wild meals are meals that are typically found in long IMU
recordings (significantly longer than the recordings used in Section
\ref{sec:algorithm}) obtained by participants during their every day
life activities. In this section we present a method for effectively
detecting the start and end time-points of in-the-wild meals using the
trained end-to-end network from Section \ref{sec:algorithm} coupled
with signal processing algorithms.

\subsection{End point detection algorithm}
\label{sec:algo_wild}

Given the $M\times 6$ observation matrix $\mathbf{R}$ that represents
an in-the-wild recording and the trained end-to-end network, the first
step is to obtain the set of detected bites $\mathcal{B}$ with
cardinality $\vert \mathcal{B} \vert = L$. The next step is to
construct the timeseries $s(n)$ with $n=1, \ldots, N$, that spans the
entire duration of the recording and is equal to one only at the
moments where a bite is detected and zero everywhere else. Formally:

\begin{equation}
  \label{step1}
    s(n)=
    \begin{cases}
      1\text{,} & \text{if }  n=b\cdot \frac{f_s}{4} \quad \forall \; b \in \mathcal{B} \\
      0\text{,} &\text{otherwise}
    \end{cases}
\end{equation}

The timeseries $s(n)$ is then convolved with a \textit{Gaussian
  filter} of length and standard deviation equal to $\frac{f_{s} \cdot
  240}{4}$ and $\frac{f_{s} \cdot 45}{4}$ samples. The latter is
important as it smooths $s(n)$ and \textit{closes} the gaps between
groups of bites that are close to each other, similar to a
morphological closing operation, which is a phenomenon present in long
meals.

We continue by replacing with zeros the elements of $s(n)$ that are
below a threshold $\lambda_{\mathbf{s}}$ and with ones the elements
that are above it; thus creating contiguous regions of fixed
magnitude. We selected threshold $\lambda_{\mathbf{s}}$ to be equal to
$5\times10^{-4}$. The selected $\lambda_{\mathbf{s}}$ value is high
enough to reject \textit{single}, isolated, peaked in the smoothed
$s(n)$ series. That simple thresholding scheme acts as a way of
filtering out sparse, false positive detections that may appear
throughout the day.

In order to extract the left-most and right-most edges of the
contiguous regions found in $s(n)$ we first convolve the thresholded
series $s(n)$ with an $1$D edge detector (or differentiation filter)
$\mathbf{h}$ with a sidelobe length equal to $f_{s}/4$ samples,
constructed as: $\mathbf{h}
=[1,2,\ldots,\frac{f_{s}}{4},0,-\frac{f_{s}}{4},\ldots,-2,-1]$. The
result of this process is the series $d(n)$ of length $N$. The final
edges of the regions in $s(n)$ are obtained by performing a local
maxima search in $|d(n)|$.

The first crude estimate of meal end-points is obtained by pairing the
consecutive peaks found in $|d(n)|$ without overlapping between pairs,
i.e. the first with the second, the third with the forth, etc. The
outcome is the set of intervals $\mathcal{Q} = \{q_{1},\ldots,q_{V}\}$
where each $q_{i}=[t^{l}_{i},t^{r}_{i}]$ with $i=1,\ldots,V$, contains
the timestamps of the left-most ($t^{l}_{i}$) and right-most
($t^{r}_{i}$) edges of the meal estimates. Next, we iteretively merge
$q_{i}$ intervals that are within $180$ seconds of each other. We
selected the merge threshold to be equal to $180$ as it adequately
merged nearby meal regions and has marginal effect in the performance
of the approach. Finally, all $q_{i}$ intervals with duration less
that $180$ seconds are rejected. Figure \ref{fig:meal_figs} depicts
the steps of the meal detection process.

\section{Experiments \& Evaluation}
\label{sec:experimentsevaluation}
\subsection{Datasets}
\label{sec:dataset}
In our experiments we make use of three datasets, namely FIC, FreeFIC
and FreeFIC held-out. All datasets contain the triaxial acceleration
and orientation velocity ($6$ DoF) signals originating from a
commercial smartwatch. However, the FIC dataset differs from the other
two regarding the context and the duration of the recordings. More
specifically, recordings in FIC focus on in-meal behavior and
therefore only contain meals, i.e. the recordings start and end when a
meal starts and ends. On the other hand, FreeFIC and FreeFIC held-out
recordings are \textit{significantly} longer in duration (average FIC
recording duration is $703$ seconds as opposed to $17,398$ seconds in
FreeFIC and $21,234$ seconds in FreeFIC held-out) as they aim to
detect eating episodes among the every day life activities of
participants. Details on the datasets are as follows.
\subsubsection{The FIC dataset}
\label{sec:fic_data}
FIC includes the inertial data from $21$ meal sessions belonging to
$12$ unique subjects, recorded in the restaurant of Aristotle
university of Thessaloniki. We used the Microsoft Band $2^{\text{TM}}$
for capturing ten out of the twenty-one meals and the Sony Smartwatch
$2^{\text{TM}}$ for the rest.

Prior to each meal recording, a GoPro$^{\text{TM}}$ Hero $5$ camera
was already set to the participant's table. The camera was mounted on
a small $23$ cm in height tripod with the ability to capture the
participants torso, face and food tray simultaneously. We used the
video streams to annotate the \textit{start} and \textit{end} moments
of each food intake cycle (i.e. bite) event that would serve as ground
truth. No special instructions were given to the subjects, other than
clapping their hands once before starting their meal and once after
they were done. This is important since the clapping motion has a very
distinctive footprint on the accelerometer signal (especially in the
magnitude series that for a specific moment $m$ is defined as
$m_{a}(m)=\sqrt{(a_{x}^{2}(m) + a_{y}^{2}(m) + a_{z}^{2}(m))}$) which
enables the synchronization between the inertial data and the video
stream. The participants were free to select the starter, salad, main
and desert of their preference, thus creating a diverse set of food
types such as, but not limited to, different kinds of meat or fish
(such as chicken legs, lamb, pork, sea bass), different kinds of
garnish (such as potatoes or rice), pasta, cooked vegetables (such as
green beans or peas), soups and various types of salads (such as
Greek-style salad, cauliflower or cabbage). Finally, FIC dataset does
not contain liquid intake instances or eating without the fork, spoon
and knife.

Given the video stream and inertial data for a meal we can create each
meal's label series $\mathbf{y}$.

\begin{equation}
  \label{eq:label_fic}
    y_{i}=
    \begin{cases}
      +1\text{,} & \text{if }  t^{gt}_{j}-\epsilon \leq t^{y}_{i} \leq t^{gt}_{j}+\epsilon\\
      -1\text{,} &\text{otherwise}
    \end{cases}
\end{equation}

Where $y_{i}$ is the binary label associated with the $i$-th windowed
data frame $\mathbf{X}_{i}$ (Section \ref{sec:end2end}), $t^{gt}_{j}$
represents the timestamp \textit{at the end} of the $j$-th bite event
and $t^{y}_{i}$ is the timestamp associated with $y_{i}$. In our
experiments we set $\epsilon$ equal to $0.1$ seconds. Figure
\ref{fig:learning_labels} a) depicts the labeling process for
recordings in FIC. Table \ref{tab:fic_statistics} provides the
statistics for the FIC dataset.

Sliding window parameters $w_l$ and $w_s$ for extracting training data
from the FIC dataset (Section \ref{sec:end2end}) were set to $5 \cdot
f_s$ and $0.05 \cdot f_s$ samples, or $5$ and $0.05$ seconds
respectively. We selected this value for $w_l$ as this window length
approximates the mean food intake cycle duration in our dataset (Table
\ref{tab:fic_statistics}).

\begin{table}[t]
  \centering
  \caption{FIC dataset statistics}
  \label{tab:fic_statistics}
  \begin{tabular}{l| c c}
    \toprule
    & \textbf{Meal sessions} & \textbf{Food Intake Cycles} \\ \midrule
    \textbf{\#} & $21$ &  $1,332$\\
    \textbf{Mean (sec)}  & $703.56$ & $4.52$ \\
    \textbf{Std (sec)}  & $186.18$& $3.22$\\
    \textbf{Median (sec)} & $717.88$ & $3.55$ \\
    \textbf{Total (sec)} & $14,774.80$ & $6,023.07$\\
    \textbf{Total (hours)} & $4.10$ & $1.67$\\ \midrule
    \textbf{Participants} & \multicolumn{2}{c}{$12$} \\
    \textbf{Ratio total/bites} & \multicolumn{2}{c}{$2.4$}\\ \bottomrule
    \end{tabular}
\end{table}

\subsubsection{The FreeFIC dataset}
\label{sec:freefic_data}
FreeFIC includes $16$ in-the-wild sessions that belong to $6$ unique
subjects. Four out of the six subjects in FreeFIC also participated in
FIC. Participants were instructed to wear the smartwatch to the hand
of their preference well ahead before any meal and continue to wear it
throughout the day until the battery is depleted. We used the Huawei
Watch $2^\text{TM}$ for $6$ out of the $16$ recordings and the Mobvoi
TicWatch$^\text{TM}$ for the rest. Similar to the FIC dataset, FreeFIC
only contains recordings were the subjects consumed their meals using
the fork and/or the spoon. In addition, we followed a self-report
labeling model, meaning that the ground truth is provided from the
participant by documenting the start and end moments of their meals to
the best of their abilities as well as the hand they wear the
smartwatch on.

Using the self-reports and the inertial data we can construct the
label series $\mathbf{y}$. More formally:

\begin{equation}
  \label{eq:label_freefic}
    y_{i}=
    \begin{cases}
      N/A\text{,} & \text{if }  t^{s}_{j} \leq t^{y}_{i} \leq t^{e}_{j} \\
      -1\text{,} &\text{otherwise}
    \end{cases}
\end{equation}

\begin{figure}[t]
  \centering
  \begin{tabular}{@{}c@{}}
    \includegraphics[width=1\linewidth]{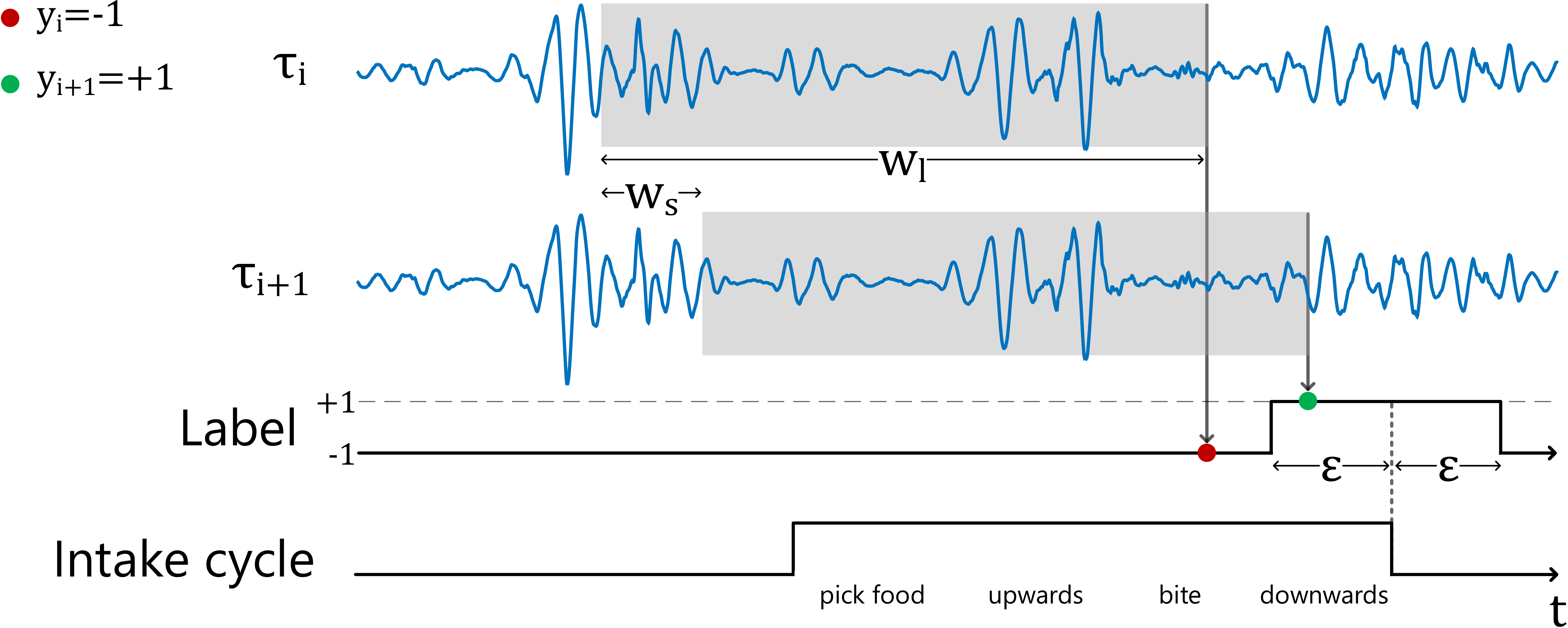} \\[\abovecaptionskip]
    \small (a) 
  \end{tabular}

  \vspace{\floatsep}

  \begin{tabular}{@{}c@{}}
    \includegraphics[width=1\linewidth]{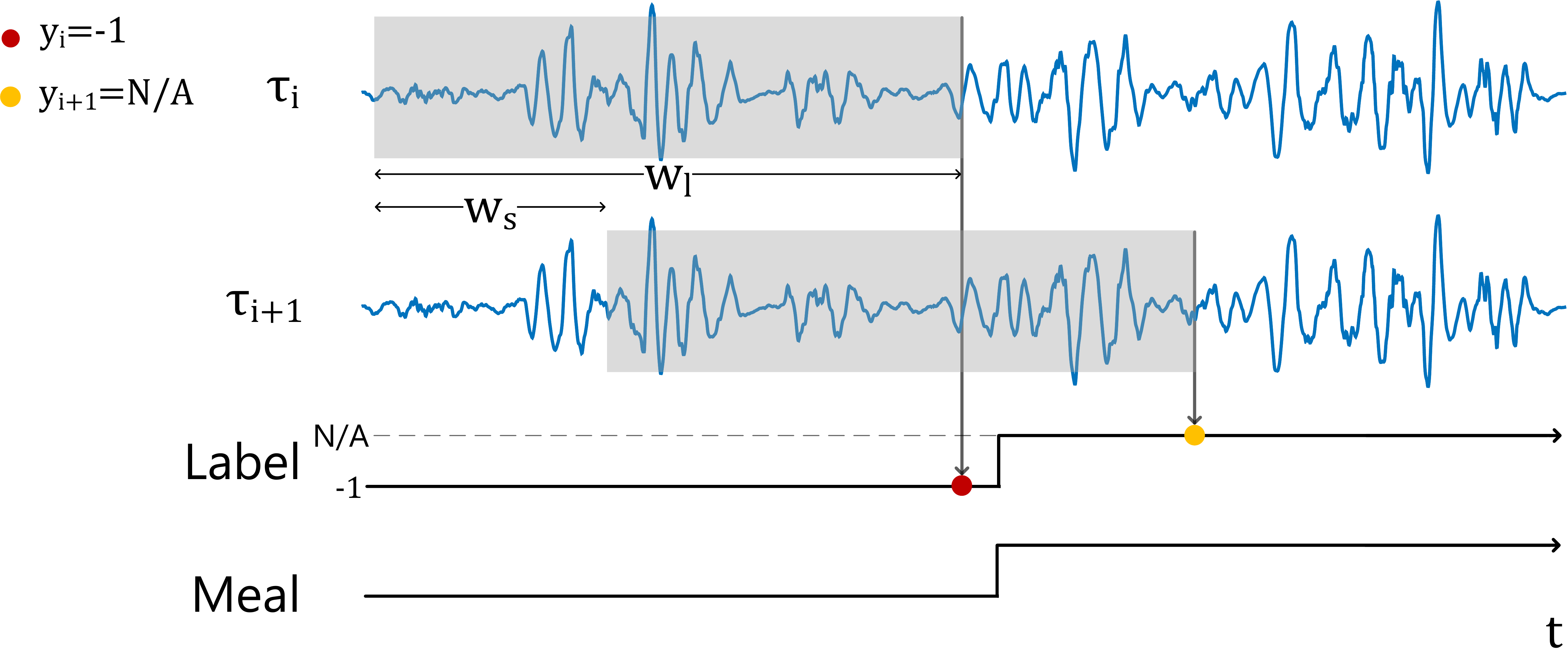} \\[\abovecaptionskip]
    \small (b) 
  \end{tabular}

  \caption{Example label extraction when $\mathbf{R}$ represents a) a
    meal recording or b) a free-living recording including one or
    more meals. Both subfigures depict the labeling process of two
    consecutive sliding windows at moments $\tau_{i}$ and
    $\tau_{i+1}$. The sliding window is illustrated by the shaded
    background while $w_l$ and $w_s$ represent it's length and
    step. In the lower part of the subfigure a), for every moment $t$
    the label signal is \textit{positive} at the end of a food intake
    cycle ($\pm\epsilon$) and \textit{negative} everywhere else. On
    the other hand for subfigure b), the label signal is negative
    everywhere except during the meals where it is \textit{not
      applicable} or N/A. In both cases the label assigned to each
    window corresponds to the label at it's right edge. For subfigure
    a), the window at $\tau_i$ is assigned a negative label while the
    window at $\tau_{i+1}$ a positive. Accordingly for subfigure b),
    the window at $\tau_i$ is also assigned a negative while the
    window at $\tau_{i+1}$ a N/A label.}\label{fig:learning_labels}
\end{figure}

Where $t^{s}_{j}$ and $t^{e}_{j}$ represent the start and end
timestamps of the $j$-th meal in a free-living recording. Variables
$y_{i}$ and $t^{y}_{i}$ are the same as in FIC. We use the label
characterization \textit{Not Applicable} (N/A) to signify that we
cannot consider this label either as positive or as negative due to
the high uncertainty of when the subject is performing a bite during a
meal. Figure \ref{fig:learning_labels} b) depicts the labeling process
for recordings in FreeFIC. Detailed statistics about FreeFIC are
provided in Table \ref{tab:freefic_statistics}.

For the extraction of training samples from FreeFIC we used the same
sliding window length $w_l$ but significantly increased the step
$w_s$, from $0.05$ to $1$ \textit{sec}. The reason behind the
$\times20$ increase is to avoid oversegmenting the significantly
longer FreeFIC recordings, which can solely yield negative training
examples.

\begin{table}[t]
  \centering
  \caption{FreeFIC dataset statistics}
  \label{tab:freefic_statistics}
  \begin{tabular}{l| c c}
    \toprule
    & \textbf{In-the-wild sessions} & \textbf{Meals} \\ \midrule
    \textbf{\#} & $16$ &  $17$\\
    \textbf{Mean (sec)} & $17,398$ & $1,148$ \\
    \textbf{Std (sec)} & $4,884$ & $502$ \\
    \textbf{Median (sec)} & $16,489$ & $1,065$  \\
    \textbf{Total (sec)} & $278,378$ & $19,520$\\
    \textbf{Total (hours)} & $77.32$ & $5.42$\\ \midrule
    \textbf{Participants} & \multicolumn{2}{c}{$6$} \\
    \textbf{Ratio total/meal} & \multicolumn{2}{c}{$14.2$}\\ \bottomrule

    \end{tabular}
\end{table}

\subsubsection{The FreeFIC held-out dataset}
\label{sec:freefic_heldout_data}
The FreeFIC held-out dataset includes $6$ in-the-wild sessions that
belong to $6$ new unique subjects (one recording per participant). All
recordings were captured using the Mobvoi TicWatch$^\text{TM}$
smartwatch. The collection protocol and instructions provided to the
participants were the same as the FreeFIC dataset (Section
\ref{sec:freefic_data}. However, in contrast to FreeFIC, no training
examples were extracted from this held-out set. The FreeFIC held-out
dataset was collected solely for testing purposes. It should also be
emphasized that there is \textit{no overlap} between the subjects in
this held-out set and the FreeFIC/FIC datasets. Detailed statistics
about FreeFIC held-out are presented in Table
\ref{tab:freeficplus_statistics}.


\begin{table}[t]
  \centering
  \caption{FreeFIC held-out dataset statistics}
  \label{tab:freeficplus_statistics}
  \begin{tabular}{l| c c}
    \toprule
    & \textbf{In-the-wild sessions} & \textbf{Meals} \\ \midrule
    \textbf{\#} & $6$ &  $6$\\
    \textbf{Mean (sec)} & $21,234$ & $960$ \\
    \textbf{Std (sec)} & $1,843$ & $103$ \\
    \textbf{Median (sec)} & $21,876$ & $960$  \\
    \textbf{Total (sec)} & $127,409$ & $5,760$\\
    \textbf{Total (hours)} & $35.39$ & $1.6$\\ \midrule
    \textbf{Participants} & \multicolumn{2}{c}{$6$} \\ 
    \textbf{Ratio total/meal} & \multicolumn{2}{c}{$22.1$}\\ \bottomrule
    \end{tabular}
\end{table}

\subsection{Experiments \& evaluation methodology}
\label{sec:expeval}

\subsubsection{Food intake cycle detection experiments - EXI}
\label{sec:exi}
The first series of experiments deals with measuring the performance of
the end-to-end approach towards the detection of food intake cycles
(i.e. bites) \textit{during} the course of a meal. Specifically, we
are interested in the cross-subject performance, or in other words how
the method performs on meals that belong to unseen subjects. To
achieve this, we used the positive and negative examples from the FIC
dataset, $(\mathbf{X}_{i},y_{i})$ $\in$ FIC $\forall$ $i$ that satisfy
$y_{i}=\pm 1$, as well as the negative examples from the FreeFIC
dataset, $(\mathbf{X}_{i},y_{i})$ $\in$ FreeFIC $\forall$ $i$ that
satisfy $y_{i}=-1$, to train our end-to-end network in a
Leave-One-Subject-Out (LOSO) fashion. We discarded every
$(\mathbf{X}_{i},y_{i})$ $\in$ FreeFIC pair where the respective
$y_{i}$ had an N/A value (i.e. during a meal). At each LOSO step we
iteretively exclude the recordings (both meal \textit{and} in-the-wild
sessions) of a single subject from the training process and use the
meal sessions of the left-out subject to perform evaluation.

In addition, we extended EXI to include a series of experiments that
aim to study the effects of synthetically augmenting the training set,
according to the methodology presented in Section \ref{sec:end2end},
and how this can influence the food intake detection performance. The
training and evaluation setup of this series of experiments is
identical to EXI.

In all cases, we measure the performance of the bite detection
algorithm by calculating the number of true positive (TP), false
positive (FP) and false negative (FN) samples. It should be noted that
the proposed in-meal bite detection evaluation scheme as well as the
one proposed by Dong \textit{et al.} in \cite{dong2012new} cannot
measure the number of True Negatives (TNs). Subsequently we can
calculate the precision, recall and F$1$ metrics defined as
$\text{Prec}=\frac{\text{TP}}{\text{TP}+\text{FP}}$,
$\text{Rec}=\frac{\text{TP}}{\text{TP}+\text{FN}}$ and
$\text{F1}=\frac{2 \cdot \text{Prec} \cdot
  \text{Rec}}{\text{Prec}+\text{Rec}}$, respectively. In detail, given
the set of detected bite moments, $\mathcal{B}$, and the set of
in-meal ground truth intervals $\mathcal{G}^{m} = \{[t^{s}_{1},
  t^{e}_{1}],\ldots,[t^{s}_{N}, t^{e}_{N}]\}$ calculation of the above
metrics can be done in the following fashion:

\begin{itemize}
\item{If a detected bite, $b_{i} \in [t^{s}_{j},t^{e}_{j}]$ exists for
  some $i$ and $j$ then we associate the $i$-th detection to the
  $j$-th ground truth interval:}
  \begin{itemize}
  \item{If no other detection $b_k$, $k \neq i$ is associated with $j$, then it counts as a TP.}
  \item{If a detection moment $b_k$ exists that is already associated
    with the $j$-th interval, then $b_i$ counts as a FP (i.e., at most
    one detected bite is associated with each ground truth interval).}
  \end{itemize}
  \item{Detection moments that satisfy $b_{i} \not\in
    [t^{s}_{j},t^{e}_{j}]$ for all $j$ count as FPs as well.}
    \item{Any ground truth interval $[t^{s}_{j}, t^{e}_{j}]$ that
      isn't associated with a detection $b_i$ for any $i$, counts as a
      FN.}
\end{itemize}
Figure \ref{fig:eval_figs} a) illustrates the proposed metric
calculation scheme for EXI.

\subsubsection{In-the-wild meal detection experiment - EXII}
\label{sec:exii}
The second series of experiments deals with measuring the performance
of the meal detection method using in-the-wild recordings of FreeFIC
(Table \ref{tab:freefic_statistics}). Experiment EXII is tightly
linked with EXI in the sense that it makes use of the trained
end-to-end networks $\mathbf{W}^{i}$ that were produced during the
LOSO iterations of EXI, with $i=1,\ldots,S$ representing the
identifier of the left-out subject out of the total $S$ subjects that
participated in EXI, which for our experimental setup equals to $14$
(the number of unique subjects in the union of FIC and FreeFIC
datasets). In more detail, for every subject $i$ in FreeFIC we use the
trained end-to-end network $\mathbf{W}^{i}$ to produce the set of
detected bites $\mathcal{B}^{i}_{k}$ for each of the $i$-th
participant's $k=1,\ldots,K$ total in-the-wild sessions. Subsequently
each $\mathcal{B}^{i}_{k}$ is propagated to the meal detection
algorithm (Section \ref{sec:algo_wild}) in order to produce the final
set of meal estimates $\mathcal{Q}^{i}_{k}$.

Similar to the evaluation of the bite detection approach, we can
measure the effectiveness of the meal detection algorithm by
exhaustively calculating the TP, FP, FN and TN metrics. Ability to
measure TNs allows us to also calculate specificity and accuracy,
defined as $\text{Spec}=\frac{\text{TN}}{\text{TN}+\text{FP}}$ and
$\text{Acc}=\frac{\text{TP}+\text{TN}}{\text{TP}+ \text{FN} +
  \text{FP}+\text{TN}}$. Motivated by \cite{dong2014detecting}, we opt
to weight TP to TN at a ratio of $14.2$:$1$. This is due to the fact
eating activities occupy a very small portion of time in free-living
scenarios. The $14.2$ weighting ratio results from Table
\ref{tab:freefic_statistics}, by dividing the total recording duration
with the time spend during meals, specifically
$77.32/5.42=14.2$. Applying the weighting factor to the accuracy
formula leads us to the \textit{weighted accuracy} (as in Section II-C
of \cite{dong2014detecting}) defined as:

\begin{equation}
  \textrm{Acc}_{w} = \frac{\textrm{TP} \cdot
    14.2+\textrm{TN}}{(\textrm{TP}+\textrm{FN})\cdot 14.2 +
    \textrm{FP}+\textrm{TN}}
\end{equation}

\begin{figure}[t]
  \centering
  \begin{tabular}{@{}c@{}}
    \includegraphics[width=1\linewidth]{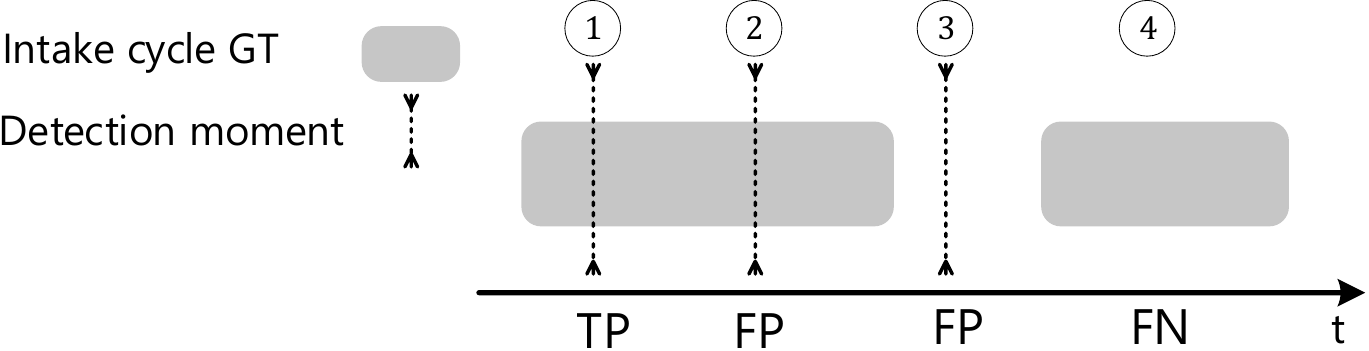} \\[\abovecaptionskip]
    \small (a) 
  \end{tabular}

  \vspace{\floatsep}

  \begin{tabular}{@{}c@{}}
    \includegraphics[width=1\linewidth]{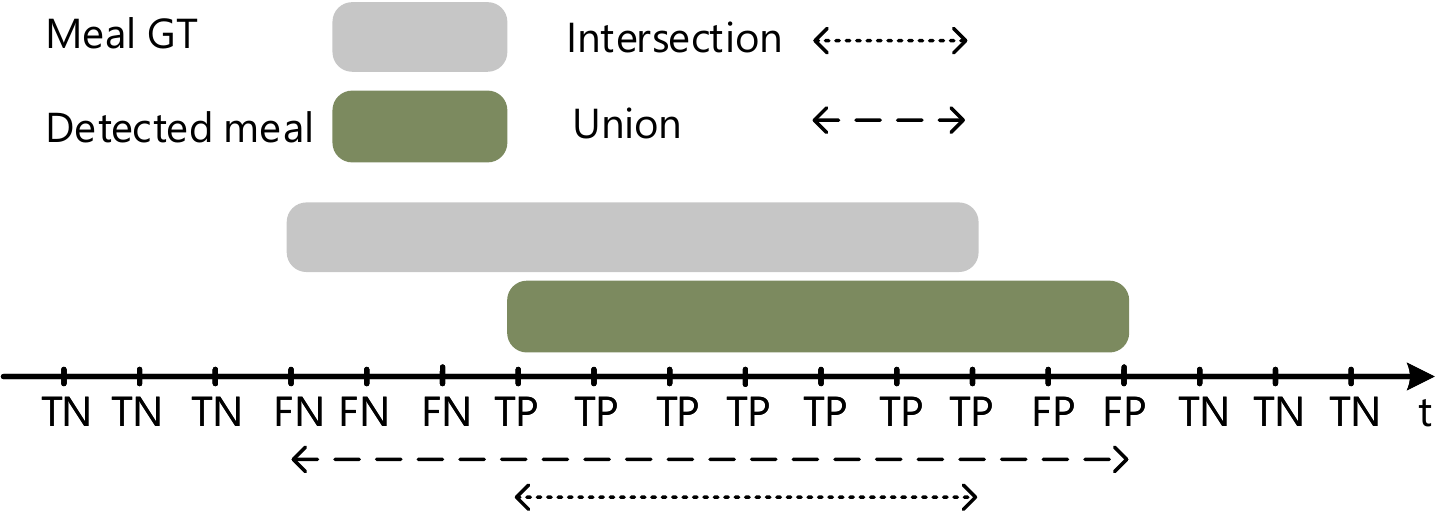} \\[\abovecaptionskip]
    \small (b) 
  \end{tabular}

  \caption{Proposed metric calculation schemes for measuring the
    effectiveness of the a) bite detection and b) meal detection
    algorithms.}\label{fig:eval_figs}
\end{figure}

In order to measure TPs, FPs, TNs and FNs, we consider the discrete
complete free-living timeline and that each point in time that
corresponds to a $q_i$ interval (i.e. in-between $t^{l}_{i}$ and
$t^{r}_{i}$) belongs to the positive class (i.e. meal) and every other
point to the negative (i.e. non-meal). Moreover, we can also assess
the meal detection effectiveness by calculating the overlap of the
detected meal intervals against the real ones using the
\textit{Jaccard Index} metric, defined as
$\mathcal{J}(\mathcal{Q},\mathcal{T})= \frac{|\mathcal{Q}\cap
  \mathcal{T}|}{|\mathcal{Q}\cup \mathcal{T}|}$, where $\mathcal{Q}$
and $\mathcal{T}$ are the estimated and true meal intervals,
respectively. Figure \ref{fig:eval_figs} b) depicts an example of how
we measure the effectiveness of the meal detection algorithm.

\subsubsection{In-the-wild held-out meal detection experiment - EXIII}
\label{sec:exiii}
The third series of experiments also deals with the detection of meals
using in-the-wild recordings. However, EXIII primarily focuses on
supporting the outcomes of the LOSO experiments (EXII, Section
\ref{sec:exii}) and evaluating the generalization ability of the
presented approach. To this end, in EXIII we use the FIC and FreeFIC
datasets (Sections \ref{sec:fic_data} and \ref{sec:freefic_data}) as
training sets and the FreeFIC held-out dataset (Section
\ref{sec:freefic_heldout_data}) as the test set (no overlapping
subjects between the FIC/FreeFIC and the FreeFIC held-out datasets).

In more detail, in EXIII we use all available data from FIC and
FreeFIC to train a \textit{single} end-to-end network $\mathbf{W}$. It
should be emphasized that network $\mathbf{W}$ was built
\textit{before} collecting the FreeFIC held-out dataset (Section
\ref{sec:freefic_heldout_data}). The network is then used to produce
the set of detected bites for each one of the $6$ in-the-wild sessions
belonging to the $6$ subjects participating in the FreeFIC held-out
set. Finally, each set of detected bites is propagated to the meal
detection algorithm in order to obtain the final set of of meal
estimates.

Performance metrics are extracted in the same fashion as in EXII, with
the exception of the accuracy weighting factor which is calculated
from Table \ref{tab:freeficplus_statistics} as $35.39/1.6=22.1$.

\subsubsection{In-the-wild meal detection using external dataset experiment - EXIV}
\label{sec:exiv}

Similar to EXII and EXIII (Sections \ref{sec:exii} and
\ref{sec:exiii}), the fourth series of experiments (EXIV) also deals
with the detection of meals using in-the-wild recordings. The
experiments performed in this section aim to to further validate the
proposed method's ability to generalize on previously unseen data,
using a publicly available external dataset with \textit{similar}
information. In more detail, we opted to use the ACE Free-living
dataset presented by Mirtchouk \textit{et al.} in
\cite{mirtchouk2017recognizing}.

The ACE Free-living dataset consists of two sub-datasets, ACE-FL and
ACE-E, and contains a total of $25$ in-the-wild recordings from $11$
subjects ($10$ recordings belonging to $5$ subjects from ACE-FL and
$15$ recordings belonging to $6$ subjects from ACE-E). Each recording
in ACE Free-living contains: 1) $9$ DoF IMU data from two commercial
smartwatches, 2) $9$ DoF IMU data from a Google Glass (only for
ACE-FL), and 3) audio information from an earbud with internal and
external microphones. All IMU data were collected at a rate of $15$
Hz. Annotation includes the start and end moments of all eating
activities, the meal type (meal, snack or drink) and the items that
were consumed. Participants could eat their food with or without
utensils as well as using either, or even both, hands.

In the scope of EXIV, the ACE Free-living dataset differs from the
FreeFIC/FreeFIC held-out datasets in terms of: 1) the IMU sampling
frequency ($15$ against $100$ Hz), 2) the number of smartwatches used
($2$ against $1$), 3) additional meal types (meals, snacks and drinks
against just meals) and 4) usage of utensils (fork, spoon and knife)
during meals since the subjects in ACE Free-living can also eat using
their bare hands.

For the experiments performed in this section we used the $6$ DoF IMU
data (3D acceleration and orientation velocity) originating from both
hands (formulated as the $M \times 6$ observation matrices
$\mathbf{R}_{l}$ and $\mathbf{R}_{r}$ for the left and right hand,
respectively), after appropriately upsampling to the frequency that is
compatible with our approach ($f_{s} = 100$ Hz). Furthermore, the EXIV
series of experiments is performed in a similar fashion as EXIII
(Section \ref{sec:exiii}). Specifically, we use all available data
from the FIC and FreeFIC datasets to train a \textit{single}
end-to-end network $\mathbf{W}$ and use ACE Free-living (all data from
both ACE-FL and ACE-E) as the held-out test set. By independently
processing $\mathbf{R}_l$ (mirrored appropriately using Equation
\ref{eq:rot}) and $\mathbf{R}_r$ using $\mathbf{W}$, we obtain the two
sets of detected bites $\mathcal{B}_l$ and $\mathcal{B}_r$. The final
meal estimates are obtained by forwarding the combined set of bite
detections $\mathcal{B}_{lr} = \mathcal{B}_l \cup \mathcal{B}_r$ to
the meal detection algorithm. In our experiments we evaluate how well
our approach generalizes on the unseen ACE Free-living dataset using
only the annotated meals as the positive eating activities (snacks and
drinks are considered negative). We also performed an experiment using
the annotated meals \textit{and} snacks as positive eating activities
(drinks are considered negative); however, it is expected a priori to
achieve reduced accuracy since the snacks category does not always
have the meal structure that we assume in our analysis.

\subsection{Results \& discussion}
\label{sec:resdisc}

Regarding the EXI series of experiments (Section \ref{sec:exi}, we
compare the performance of the proposed algorithm against four
state-of-the art methods in the same datasets. The first one is the
work of Zhang \textit{et al.}  \cite{zhang2016food}, the second one is
the work of Dong \textit{et al.} \cite{dong2012new} and the two other
works \cite{kyritsis2019modeling} and \cite{kyritsis2017food} belong
to our group. A synopsis of the aforementioned works has been provided
in Section \ref{sec:soa}. One fundamental difference in the core of
those methods is dependency on micromovements. Specifically, the works
of \cite{kyritsis2019modeling,kyritsis2017food,zhang2016food} rely on
knowledge of micromovements while the work of \cite{dong2012new} and
the method proposed in this paper are micromovement-agnostic.

Table \ref{tab:table_results_fic_our} presents the results of EXI,
using: i) the evaluation scheme explained in Section \ref{sec:exi}
(also depicted in Figure \ref{fig:eval_figs} a) and ii) the more
relaxed evaluation scheme of \cite{dong2012new}. Evaluation results
indicate that the proposed approach outperforms all other methods,
both micromovement-based and micromovement-agnostic, by achieving an
F$1$ score of $0.928$ using the proposed evaluation scheme. Our
previous micromovement-based approach presented in
\cite{kyritsis2019modeling} achieves the second highest performance
with an F$1$ score of $0.913$ under the same scheme. When switching to
the less strict evaluation scheme of Dong \textit{et al.}
\cite{dong2012new} an overall increase to F$1$ can be observed across
all experiments. In addition to the F$1$ increase, the
performance-wise ranking of the compared methods is preserved except
the first two places where the method of \cite{kyritsis2019modeling}
marginally outperforms the proposed one with an F$1$ score of $0.929$
over $0.928$.

Table \ref{tab:augmentation} shows the effects of synthetically
augmenting the dataset during the training process of the proposed
method regarding the intake cycle detection performance. Specifically,
the best performance is obtained by allowing the training data to be
rotated both around the $x$ (roll) and $z$ (yaw) axes, thus achieving
an F$1$ score of $0.923$. Solely rotating the training data around the
$x$ or the $z$ axis yield a score of $0.911$ and $0.919$,
respectively. Overall, synthetically augmenting the dataset by
rotating around $x$ and $z$ results into a considerable improvement
over not using data augmentation at all which achieves the lowest
performance of all four trials with an F$1$ score of $0.888$.

Regarding the first series of in-the-wild meal detection experiments
(EXII, Section \ref{sec:exii}), we compare the proposed approach with
the one presented by Dong \textit{et al.} \cite{dong2014detecting}. In
contrast to our bite-based meal detection method that makes the
hypothesis of the density of the detected bites being low outside the
meal intervals, the core behind \cite{dong2014detecting} is that
\textit{before} and \textit{after} every meal a period of increased
wrist motion energy is observed. At the same time the authors also
assume that the wrist motion energy \textit{during} the course of a
meal tends to be reduced. The wrist motion formula is formally defined
by the authors as:
\begin{equation}
  \label{eq:motion_formula}
E(i) = \frac{1}{W+1} \sum_{i=t-\frac{W}{2}}^{t+\frac{W}{2}} |a_{x}(i)|
+ |a_{y}(i)| +|a_{z}(i)|
  \end{equation}
where $a_{x}(i)$, $a_{y}(i)$ and $a_{z}(i)$ are the smoothed $x$, $y$
and $z$ channels of the accelerometer sensor and $W$ is the length of
the sliding window over which the wrist motion energy is calculated
for a single moment. We implemented the method of
\cite{dong2014detecting} and subsequently used a small subset of the
in-the-wild recordings to tune: i) the length of the window size (W)
and ii) the multiplication factor between the T$1$ and T$2$
thresholds, that affects the segmentation part of their
method. Throughout our experiments, we used the complete feature set
as suggested by the authors.

Furthermore, we compare our method with \textit{Density-Based Spatial
  Clustering of Applications with Noise} (DBSCAN). The motivation
behind using DBSCAN as a solution to the meal detection problem stems
from the work of Zhang \textit{et al.}  \cite{zhang2016food} where the
authors employ DBSCAN in order to detect the final \textit{bite
  moments} using the density of bite decisions produced by their
algorithm. In this experiment we transfer the application of DBSCAN
from detecting eating moments to detecting meals found in the wild. We
provide as input to DBSCAN the $\mathcal{B}_{k}^{i}$ set of bite
timestamps, for the $k$-th session of the $i$-th subject, as computed
by our proposed end-to-end network. In addition and to be on par with
our proposed meal detection approach we parameterize DBSCAN
accordingly by setting the \textit{maximum distance between samples}
(usually mentioned as the \textit{eps} parameter) to $180$ seconds. To
obtain the final meal estimates using DBSCAN we pair the two extreme
points of each formed cluster.

The first half of Table \ref{tab:meal_results} summarizes the
performance of all meal detection algorithms in a LOSO evaluation
setting using the FreeFIC set. The results initially point out that
the proposed approach outperforms all other methods by yielding an
F$1$/weighted accuracy/Jaccard Index of $0.899$/$0.953$/$0.820$. In
addition, both methods that use the distribution of the detected bites
in order to perform meal detection exhibit increased performance.

One potential reason behind the difference between the results
reported in \cite{dong2014detecting} and the ones reported here is the
nature of the sensing device (strapped phone on the wrist against
smartwatch). In addition, the authors of both \cite{dong2014detecting}
and \cite{sharma2016automatic} report that the hypothesis of vigorous
wrist movement before and after a meal can be population- and
lifestyle-specific.

Experimental results from the second series of the in-the-wild meal
detection experiments (EXIII, Section \ref{sec:exiii}) support the
results obtained from the LOSO experiments (EXII, Section
\ref{sec:exii}. The second half of Table \ref{tab:meal_results}
presents the performance of all meal detection algorithms using the
FreeFIC held-out set as described in \ref{sec:exiii}. Comparison
between the first and second halves of Table \ref{tab:meal_results}
reveal that the proposed approach achieves almost identical results
with the ones obtain in EXII; specifically, an F$1$ score of $0.896$
($0.899$ in EXII), a weighted accuracy of $0.964$ ($0.953$ in EXII),
and a Jaccard Index of $0.821$ ($0.820$ in EXII). The similarity in
the obtained results in the two in-the-wild datasets clearly indicates
that the LOSO test error estimation of EXII is accurate and that the
proposed method generalizes in previously unseen samples. This
argument is further supported by the fact that the collection of the
FreeFIC held-out set initiated after the completion of the LOSO
experiments (EXII).

Finally, Table \ref{tab:experiment_iv} showcases the obtained
in-the-wild meal detection results when using the external
publicly-available dataset presented in
\cite{mirtchouk2017recognizing} (EXIV, Section \ref{sec:exiv}). More
specifically, by using the inertial information from both hands, the
proposed meal localization method achieves a weighted accuracy of
$0.825$ when using only the meal sessions as positive periods. In
addition, when using both the meals and snacks as positive periods,
the proposed method yields a weighted accuracy of $0.788$. Results of
this experiment on $15$ Hz sampling rate data of an external dataset
further demonstrate the robustness of our method and it's ability to
generalize on previously unseen data. It should be emphasized that
apart from upsampling the ACE IMU signals (from $15$ to $100$ Hz),
this performance was achieved without retraining the method (or tuning
any parameters) to the specific dataset.

\begin{table*}[t]
  \centering
  \caption{Bite detection performance results. The table contains the
    results of EXI using the evaluation scheme described in
    \ref{sec:exi}.The $\star$ symbol in the method proposed by Dong
    \textit{et al.} is used to signify that parameter tuning was
    performed by optimization based on our proposed evaluation scheme
    (Figure \ref{fig:eval_figs} a). The numbers inside the parentheses
    represent the results obtained using the evaluation scheme
    proposed in \cite{dong2012new}.}
  \label{tab:table_results_fic_our}
  \begin{tabular}{l c c c c c c c}
    \toprule
    \textbf{Bite detection method} & \textbf{Micromovements (y/n)} & \textbf{TP} & \textbf{FP} & \textbf{FN} &  \textbf{Prec} & \textbf{Rec} & \textbf{F1} \\ \midrule  
    Proposed & n  & $1,231$ ($1,237$) & $102$ ($96$) & $101$ ($95$) & $.923$ ($.927$) & $.924$ ($.928$) & $\mathbf{.923}$ ($.928$) \\     
    Kyritsis \textit{et al.} \cite{kyritsis2019modeling}& y & $1,241.8$ ($1,263.4$) & $144.5$ ($122.9$) & $90.2$ ($68.6$) & $.895$ ($.911$) & $.932$ ($.948$)& $.913$ ($\mathbf{.929}$)\\
    Kyritsis \textit{et al.} \cite{kyritsis2017food}& y  & $1,221.5$ ($1,267.6$) & $228.4$ ($182.3$)& $110.5$ ($64.4$) & $.842$ ($.874$)& $.917$ ($.951$)& $.878$ ($.911$)\\
    Zhang \textit{et al.}\cite{zhang2016food}& y  & $944$ ($1,102$) & $431$ ($233$) & $388$ ($230$) & $.686$ ($.825$) & $.708$ ($.827$)& $.697$($.826$) \\
    Dong \textit{et al.}\cite{dong2012new}& n & $707$ ($1,190$)& $794$ ($311$) & $625$ ($142$)& $.471$ ($.792$)& $.530$ ($.893$)& $.499$ ($.840$)\\
    Dong \textit{et al.}\cite{dong2012new} $\star$ & n & $772$ ($1,214$) & $746$ ($304$)& $560$ ($118$)& $.508$ ($.799$) & $.579$ ($.911$)& $.541$ ($.851$) \\ \bottomrule       
  \end{tabular}
\end{table*}

\begin{table*}[t]
  \centering
  \caption{Effects of proposed data augmentation scheme (Section
    \ref{sec:end2end}) applied during the training of the proposed
    end-to-end network regarding the in-meal bite detection
    performance.}
  \label{tab:augmentation}
  \begin{tabular}{l c c c c c c }
    \toprule
    \textbf{Data augmentation method} & \textbf{TP} & \textbf{FP} & \textbf{FN} & \textbf{Precision} & \textbf{Recall} & \textbf{F1} \\ \midrule
    None & $1,145$ & $101$ & $187$ & $.918$ & $.859$ & $.888$ \\
    Around \textit{x} axis (roll) & $1,215$ & $120$ & $117$ & $.910$ & $.912$ & $.911$ \\
    Around \textit{z} axis (yaw) & $1,210$ & $91$ & $122$ & $.930$ & $.908$ & $.919$ \\
    Around \textit{x} \& \textit{z} axes & $1,231$ & $102$ & $101$ & $.923$ & $.924$ & $\mathbf{.923}$ \\ \bottomrule
    \end{tabular}
\end{table*}

\begin{table*}[t]
  \centering
  \caption{In-the-wild meal start/end point detection results. The
    table presents the results of experiments EXII (first half) and
    EXIII (second half) using the proposed meal detection approach,
    DBSCAN and the one presented in \cite{dong2014detecting}.}
  \label{tab:meal_results}
  \begin{tabular}{l| l c c c c c c c }
    \toprule
    \textbf{Experiment} & \textbf{Meal detection method} &  \textbf{Prec} & \textbf{Rec} & \textbf{Spec} & \textbf{F1} & \textbf{Accuracy} & \textbf{Weighted Accuracy} & $\mathcal{J}$ \textbf{Index} \\ \midrule
    \multirow{3}{*}{EXII (LOSO)} & Proposed &$.880$ & $.919$&$.990$ &$.899$ &$.985$ & $\mathbf{.953}$&$.820$ \\
    & DBSCAN & $.838$ & $.895$ & $.986$ & $.865$ & $.980$ & $.939$ & $.752$ \\
    & Dong \textit{et al.} \cite{dong2014detecting} & $.323$  &$.525$  & $.919$ &$.400$ &$.892$  &$.545$ &$.255$ \\ \midrule
    
    \multirow{3}{*}{EXIII (Held-out set)} & Proposed &$.858$ & $.937$&$.992$ &$.896$ &$.990$ & $\mathbf{.964}$&$.821$ \\
    & DBSCAN & $.774$ & $.779$ & $.989$ & $.776$ & $.979$ & $.882$ & $.681$ \\
    & Dong \textit{et al.} \cite{dong2014detecting} & $.105$  &$.697$  & $.714$ &$.182$ &$.717$  &$.182$ &$.089$ \\ \bottomrule
  \end{tabular}
\end{table*}

\begin{table*}[t]
  \centering
  \caption{In-the-wild meal start/end point detection results obtained
    using the proposed method and the ACE Free-living dataset as the
    held-out test set (EXIV, Section \ref{sec:exiv}). Accuracy is
    weighted using a factor of $15.12$ and $12.02$ for the first
    (meals only) and the second (meals \& snacks) experiment,
    respectively.}
  \label{tab:experiment_iv}
  \begin{tabular}{l c c c c c c c}
    \toprule
    \textbf{Positive periods} & \textbf{Prec} & \textbf{Rec} & \textbf{Spec} & \textbf{F1} & \textbf{Accuracy} & \textbf{Weighted Accuracy} & $\mathcal{J}$ \textbf{Index} \\ \midrule
    Meals only  & $.397$ & $.710$ & $.934$  & $.509$ & $.921$ & $.825$ & $.346$ \\
    Meals \& snacks & $.457$ & $.633$ & $.939$ & $.531$ & $.917$ & $.788$ & $.377$ \\ \bottomrule
    \end{tabular}
\end{table*}

\section{Limitations}
\label{sec:limitations}

One of the limitations of the presented bite detection approach
(Section \ref{sec:algorithm}) is the unpredictable behavior of the
algorithm when the participant performs drinking gestures or eats
without using the fork or the spoon, e.g., using bare hands or a pair
of chopsticks. This unpredictable behavior stems from the fact that
such examples were not introduced appropriately to the network during
training. For example, liquid intake episodes that appear outside of
meals in the in-the-wild recordings of FreeFIC dataset are considered
as \textit{negative} samples. The presented work is focused on meals
centered around the use of the fork and/or the spoon.

A limitation of our evaluation approach is the inability to measure
the in-meal bite detection performance using the in-the-wild
recordings of FreeFIC and FreeFIC held-out. Despite the promising
in-meal bite detection results (EXI, first row of Table
\ref{tab:table_results_fic_our}), the appearance of single, isolated,
bite detections outside of meals in the in-the-wild recordings can be
characterized as over-estimation. However, this is handled by the
thresholding scheme proposed in Section \ref{sec:algo_wild}.
In addition, the collection protocol lacks obtaining information other
than the timestamps of the start and end moments of the meals
throughout the day. For example, in the work presented in
\cite{mirtchouk2017recognizing} the participants also took pictures
before eating their meals, thus allowing for more complex types of
future analysis.

A technical limitation of the proposed approach is the ability to be
executed in-real time. While to this day smartwatch processing
capabilities remain low, modern smartphones have begun to offer
on-chip AI support (e.g. the Huawei Ascend 910), which can provide
real time performance in cases where the models are shallow. The
proposed model consists of $160,000$ parameters which suggests that
the real-time feedforwarding small batches of data to the mobile
device is a realistic scenario. Another technical limitation regarding
the recording of IMU signals throughout the day is the high battery
consumption when capturing the gyroscope sensor. However, we have not
explored methods for overcoming either of the two technical
limitations yet. Instead, in this paper we investigate the feasibility
of an automated and objective eating behavior monitoring approach
where the collected signals are transmitted and are processed
remotely.

\section{Conclusions and future work}
\label{sec:conclusions}

In this paper, a complete framework towards the in-the-wild modeling
of eating behavior using the IMU signals from a commercial smartwatch
has been presented. The proposed framework consists of two
parts. Initially, we follow a representation learning approach that
includes an end-to-end NN with both convolutional and recurrent layers
to detect bite events. Next we use the distribution of detected bites
throughout the day to temporally localize meals by detecting their
start and end points, using signal processing algorithms.

Experimental results using $12$ subjects for the LOSO in-meal bite
detection experiments (EXI) as well as $12$ subjects for the meal
localization experiments ($6$ evaluated in a LOSO fashion in EXII and
$6$ evaluated in a held-out fashion in EXIII, without overlap between
the two populations) showcase the high potential of the proposed
approach. In more detail, we initially perform LOSO evaluation using
our publicly available FIC and FreeFIC datasets where the proposed
framework outperforms other state-of-the-art methods found in the
recent literature ($0.923$ F$1$ score for in-meal bite detection and
$0.820$/$0.953$ Jaccard Index/weighted accuracy for the temporal
localization of in-the-wild meals). In addition to the LOSO
experiments, we also perform evaluation regarding the detection of
meal start/end points using the FreeFIC held-out set where our
proposed approach achieves similar results with the LOSO experiments
($0.821$/$0.964$ Jaccard Index/weighted accuracy). Finally, the
proposed method's ability to generalize on unseen data is further
validated (by achieving $0.346$/$0.825$ Jaccard Index/weighted
accuracy) by using an external publicly-available dataset that
contains in-the-wild sessions. Overall, the results reported in
\ref{sec:resdisc} are promising and point out the potential of the
framework towards monitoring and modeling eating behavior.

Future work includes increasing the size of the in-the-wild datasets
by recruiting additional subjects as well as increasing the length of
the recordings by solving the battery issue that arises when recording
the accelerometer and gyroscope at a high sampling rate. Investigation
of drinking and/or eating without utensils gestures is also one
potential direction of the current work. Additionally, we plan to
extend our work by including additional cues, e.g., visual, with the
aim of further increasing the bite detection performance. Finally, in
an ongoing study we attempt to correlate in-meal indicators that
derive from bite detections, with bradykinesia and parkinsonian tremor
using data from healthy controls, early and advanced Parkinson's
Disease (PD) patients.

\section*{Acknowledgments}
The work leading to these results has received funding from the EU
Commission under Grant Agreement No. 727688 (http://bigoprogram.eu,
H2020).


\bibliographystyle{IEEEtran}
\bibliography{IEEEabrv,paper}

\begin{thebibliography}{10}
\providecommand{\url}[1]{#1}
\csname url@samestyle\endcsname
\providecommand{\newblock}{\relax}
\providecommand{\bibinfo}[2]{#2}
\providecommand{\BIBentrySTDinterwordspacing}{\spaceskip=0pt\relax}
\providecommand{\BIBentryALTinterwordstretchfactor}{4}
\providecommand{\BIBentryALTinterwordspacing}{\spaceskip=\fontdimen2\font plus
\BIBentryALTinterwordstretchfactor\fontdimen3\font minus
  \fontdimen4\font\relax}
\providecommand{\BIBforeignlanguage}[2]{{%
\expandafter\ifx\csname l@#1\endcsname\relax
\typeout{** WARNING: IEEEtran.bst: No hyphenation pattern has been}%
\typeout{** loaded for the language `#1'. Using the pattern for}%
\typeout{** the default language instead.}%
\else
\language=\csname l@#1\endcsname
\fi
#2}}
\providecommand{\BIBdecl}{\relax}
\BIBdecl

\bibitem{world2000obesity}
W.~H. Organization, \emph{Obesity: preventing and managing the global
  epidemic}.\hskip 1em plus 0.5em minus 0.4em\relax World Health Organization,
  2000, no. 894.

\bibitem{caballero2007global}
B.~Caballero, ``{The Global Epidemic of Obesity: An Overview},''
  \emph{Epidemiologic Reviews}, vol.~29, no.~1, pp. 1--5, 06 2007.

\bibitem{yang2010review}
C.-C. Yang and Y.-L. Hsu, ``A review of accelerometry-based wearable motion
  detectors for physical activity monitoring,'' \emph{Sensors}, 2010.

\bibitem{bonomi2012advances}
A.~Bonomi and K.~Westerterp, ``Advances in physical activity monitoring and
  lifestyle interventions in obesity: a review,'' \emph{International journal
  of obesity}, vol.~36, no.~2, p. 167, 2012.

\bibitem{schoeller1990accurate}
D.~A. Schoeller, ``{How Accurate Is Self-Reported Dietary Energy Intake?}''
  \emph{Nutrition Reviews}, vol.~48, no.~10, pp. 373--379, 10 1990.

\bibitem{papapanagiotou2018automatic}
V.~Papapanagiotou, C.~Diou, I.~Ioakimidis, P.~Sodersten, and A.~Delopoulos,
  ``Automatic analysis of food intake and meal microstructure based on
  continuous weight measurements,'' \emph{IEEE Journal of Biomedical and Health
  Informatics}, pp. 1--1, 2018.

\bibitem{korotitsch1999overview}
W.~J. Korotitsch and R.~O. Nelson-Gray, ``An overview of self-monitoring
  research in assessment and treatment.'' \emph{Psychological Assessment},
  vol.~11, no.~4, p. 415, 1999.

\bibitem{vu2017wearable}
T.~Vu \emph{et~al.}, ``Wearable food intake monitoring technologies: A
  comprehensive review,'' \emph{Computers}, vol.~6, no.~1, 2017.

\bibitem{zhang2016food}
S.~Zhang \emph{et~al.}, ``Food watch: Detecting and characterizing eating
  episodes through feeding gestures,'' ser. BodyNets '16, 2016.

\bibitem{kyritsis2019modeling}
K.~{Kyritsis}, C.~{Diou}, and A.~{Delopoulos}, ``Modeling wrist micromovements
  to measure in-meal eating behavior from inertial sensor data,'' \emph{IEEE
  Journal of Biomedical and Health Informatics}, pp. 1--1, 2019.

\bibitem{dong2012new}
Y.~Dong \emph{et~al.}, ``A new method for measuring meal intake in humans via
  automated wrist motion tracking,'' \emph{Applied psychophysiology and
  biofeedback}, vol.~37, no.~3, 2012.

\bibitem{mertes2018measuring}
G.~Mertes \emph{et~al.}, ``Measuring weight and location of individual bites
  using a sensor augmented smart plate,'' in \emph{Engineering in Medicine and
  Biology Society (EMBC)}, 2018.

\bibitem{kalantarian2015audio}
H.~Kalantarian and M.~Sarrafzadeh, ``Audio-based detection and evaluation of
  eating behavior using the smartwatch platform,'' \emph{Computers in Biology
  and Medicine}, vol.~65, pp. 1 -- 9, 2015.

\bibitem{passler2012food}
S.~Pä{\ss}ler, M.~Wolff, and W.-J. Fischer, ``Food intake monitoring: an
  acoustical approach to automated food intake activity detection and
  classification of consumed food,'' \emph{Physiological Measurement}, vol.~33,
  no.~6, pp. 1073--1093, may 2012.

\bibitem{anthimopoulosfood2014}
M.~M. Anthimopoulos \emph{et~al.}, ``A food recognition system for diabetic
  patients based on an optimized bag-of-features model,'' \emph{IEEE Journal of
  Biomedical and Health Informatics}, vol.~18, no.~4, 2014.

\bibitem{zhu2011multilevel}
F.~Zhu, M.~Bosch, N.~Khanna, C.~J. Boushey, and E.~J. Delp, ``Multilevel
  segmentation for food classification in dietary assessment,'' in \emph{2011
  7th International Symposium on Image and Signal Processing and Analysis
  (ISPA)}, Sept 2011, pp. 337--342.

\bibitem{kawano2013real}
Y.~Kawano and K.~Yanai, ``Real-time mobile food recognition system,'' in
  \emph{The IEEE Conference on Computer Vision and Pattern Recognition (CVPR)
  Workshops}, June 2013.

\bibitem{kong2012dietcam}
F.~Kong and J.~Tan, ``Dietcam: Automatic dietary assessment with mobile camera
  phones,'' \emph{Pervasive and Mobile Computing}, vol.~8, 2012.

\bibitem{papapanagiotou2017chewing}
V.~Papapanagiotou, C.~Diou, and A.~Delopoulos, ``Chewing detection from an
  in-ear microphone using convolutional neural networks,'' in \emph{Engineering
  in Medicine and Biology Society (EMBC)}, July 2017.

\bibitem{pabler2011acoustical}
S.~Päßler and W.~J. Fischer, ``Acoustical method for objective food intake
  monitoring using a wearable sensor system,'' in \emph{2011 5th International
  Conference on Pervasive Computing Technologies for Healthcare
  (PervasiveHealth) and Workshops}, May 2011, pp. 266--269.

\bibitem{thomaz2015practical}
E.~Thomaz, I.~Essa, and G.~D. Abowd, ``A practical approach for recognizing
  eating moments with wrist-mounted inertial sensing,'' in \emph{Proceedings of
  the 2015 ACM International Joint Conference on Pervasive and Ubiquitous
  Computing}.\hskip 1em plus 0.5em minus 0.4em\relax ACM, 2015.

\bibitem{dong2014detecting}
Y.~{Dong} \emph{et~al.}, ``Detecting periods of eating during free-living by
  tracking wrist motion,'' \emph{IEEE Journal of Biomedical and Health
  Informatics}, vol.~18, no.~4, pp. 1253--1260, July 2014.

\bibitem{gomes2019real}
D.~Gomes and I.~Sousa, ``Real-time drink trigger detection in free-living
  conditions using inertial sensors,'' \emph{Sensors}, vol.~19, no.~9, 2019.

\bibitem{ortega2019eating}
D.~Ortega~Anderez, A.~Lotfi, and A.~Pourabdollah, ``Eating and drinking gesture
  spotting and recognition using a novel adaptive segmentation technique and a
  gesture discrepancy measure,'' \emph{Expert Systems with Applications}, 2019.

\bibitem{papapanagiotou2017novel}
V.~Papapanagiotou, C.~Diou, L.~Zhou, J.~van~den Boer, M.~Mars, and
  A.~Delopoulos, ``A novel chewing detection system based on ppg, audio, and
  accelerometry,'' \emph{IEEE Journal of Biomedical and Health Informatics},
  vol.~21, no.~3, pp. 607--618, May 2017.

\bibitem{bi2017familylog}
{Chongguang Bi} \emph{et~al.}, ``Familylog: A mobile system for monitoring
  family mealtime activities,'' in \emph{2017 IEEE International Conference on
  Pervasive Computing and Communications (PerCom)}, 2017.

\bibitem{mirtchouk2017recognizing}
M.~Mirtchouk \emph{et~al.}, ``Recognizing eating from body-worn sensors:
  combining free-living and laboratory data,'' \emph{Proceedings of the ACM on
  Interactive, Mobile, Wearable and Ubiquitous Technologies}, vol.~1, no.~3,
  2017.

\bibitem{zhang2018monitoring}
R.~{Zhang} and O.~{Amft}, ``Monitoring chewing and eating in free-living using
  smart eyeglasses,'' \emph{IEEE Journal of Biomedical and Health Informatics},
  vol.~22, no.~1, pp. 23--32, Jan 2018.

\bibitem{fontana2014automatic}
J.~M. {Fontana}, M.~{Farooq}, and E.~{Sazonov}, ``Automatic ingestion monitor:
  A novel wearable device for monitoring of ingestive behavior,'' \emph{IEEE
  Transactions on Biomedical Engineering}, vol.~61, no.~6, 2014.

\bibitem{mirtchouk2019automated}
M.~Mirtchouk \emph{et~al.}, ``Automated estimation of food type from body-worn
  audio and motion sensors in free-living environments,'' in \emph{Machine
  Learning for Healthcare Conference}, 2019, pp. 641--662.

\bibitem{heydarian2019assessing}
H.~Heydarian \emph{et~al.}, ``Assessing eating behaviour using upper limb
  mounted motion sensors: A systematic review,'' \emph{Nutrients}, 2019.

\bibitem{shen2017assessing}
Y.~Shen, J.~Salley, E.~Muth, and A.~Hoover, ``Assessing the accuracy of a wrist
  motion tracking method for counting bites across demographic and food
  variables,'' \emph{IEEE Journal of Biomedical and Health Informatics},
  vol.~21, no.~3, pp. 599--606, May 2017.

\bibitem{kyritsis2018end}
K.~{Kyritsis}, C.~{Diou}, and A.~{Delopoulos}, ``End-to-end learning for
  measuring in-meal eating behavior from a smartwatch,'' in \emph{Engineering
  in Medicine and Biology Society (EMBC)}, July 2018.

\bibitem{kyritsis2017food}
K.~Kyritsis, C.~Diou, and A.~Delopoulos, ``Food intake detection from inertial
  sensors using {LSTM} networks,'' in \emph{International Conference on Image
  Analysis and Processing}.\hskip 1em plus 0.5em minus 0.4em\relax Springer,
  2017, pp. 411--418.

\bibitem{farooq2016detection}
M.~{Farooq} and E.~{Sazonov}, ``Detection of chewing from piezoelectric film
  sensor signals using ensemble classifiers,'' in \emph{Engineering in Medicine
  and Biology Society (EMBC)}, Aug 2016, pp. 4929--4932.

\bibitem{sharma2016automatic}
S.~{Sharma} \emph{et~al.}, ``Automatic detection of periods of eating using
  wrist motion tracking,'' in \emph{First International Conference on Connected
  Health: Applications, Systems and Engineering Technologies}, 2016.

\bibitem{kyritsis2019detecting}
K.~{Kyritsis}, C.~{Diou}, and A.~{Delopoulos}, ``Detecting meals in the wild
  using the inertial data of a typical smartwatch,'' in \emph{Engineering in
  Medicine and Biology Society (EMBC)}, 2019.

\bibitem{simonyan2014very}
K.~Simonyan and A.~Zisserman, ``Very deep convolutional networks for
  large-scale image recognition,'' \emph{CoRR}, vol. abs/1409.1556, 2014.

\bibitem{srivastava2014dropout}
N.~Srivastava \emph{et~al.}, ``Dropout: A simple way to prevent neural networks
  from overfitting,'' \emph{J. Mach. Learn. Res.}, vol.~15, no.~1, 2014.

\end{thebibliography}
\end{document}